\documentclass[preprint,1p,times]{elsarticle}

\usepackage{amsmath,amssymb,amsfonts}
\usepackage{algorithmic}
\usepackage{graphicx}
\usepackage{textcomp}
\usepackage{subfigure}
\usepackage{booktabs}
\usepackage{makecell}
\usepackage{multirow}
\usepackage {color}

\usepackage{hyperref}
\usepackage{setspace}
\usepackage{bm}
\usepackage[normalem]{ulem}
\usepackage{hyperref}
\usepackage{stfloats}
\usepackage[justification=centering]{caption}
\usepackage{setspace} 
\usepackage{setspace} 
\usepackage{array}
\usepackage{makecell}
\usepackage{changes}
\usepackage{mdframed}
\usepackage{multicol}
\usepackage{multirow}
\usepackage{framed} 
\usepackage{multicol} 
\usepackage{booktabs}
\usepackage{threeparttable}
\usepackage{adjustbox}
\usepackage[T1]{fontenc}
\usepackage[tableposition=top]{caption}
\usepackage{tabularx}
\usepackage{soul}
\usepackage{url}
\usepackage{nomencl} 
\biboptions{numbers,sort&compress}

\makeatletter
\def\ps@pprintTitle{%
 \let\@oddhead\@empty
 \let\@evenhead\@empty
 \let\@oddfoot\@empty
 \let\@evenfoot\@oddfoot
}
\makeatother

\journal{Renewable Energy}

\begin{document}

\begin{frontmatter}

\title{Collective Large-scale Wind Farm Multivariate Power Output Control Based on Hierarchical Communication Multi-Agent Proximal Policy Optimization}

\author[main,second]{Yubao~Zhang} \author[main,second]{Xin~Chen\corref{cor1}} \author[main,second]{Sumei~Gong} \author[third]{Haojie~Chen}
\cortext[cor1]{E-mail: xin.chen.nj@xjtu.edu.cn}
\address[main]{School of Electrical Engineering, Xi'an Jiaotong University, Xi'an 710054, Shaanxi, China.}
\address[second]{Center of Nanomaterials for Renewable Energy, State Key Laboratory of Electrical Insulation and Power Equipment, School of Electrical Engineering, Xi'an Jiaotong University, Xi'an 710054, Shaanxi, China.}
\address[third]{China Academy of Industrial Internet, Beijing 100016, Beijing, China.}

\begin{abstract}
Wind power is becoming an increasingly important source of renewable energy worldwide. However, wind farm power control faces significant challenges due to the high system complexity inherent in these farms. A novel communication-based multi-agent deep reinforcement learning large-scale wind farm multivariate control is proposed to handle this challenge and maximize power output. A  wind farm multivariate power model is proposed to study the influence of wind turbines (WTs) wake on power. The multivariate model includes axial induction factor, yaw angle, and tilt angle controllable variables. 
The hierarchical communication multi-agent proximal policy optimization (HCMAPPO) algorithm is proposed to coordinate the multivariate large-scale wind farm continuous controls. The large-scale wind farm is divided into multiple wind turbine aggregators (WTAs), and neighboring WTAs can exchange information through hierarchical communication to maximize the wind farm power output. Simulation results demonstrate that the proposed multivariate HCMAPPO can significantly increase wind farm power output compared to the traditional PID control, coordinated model-based predictive control, and multi-agent deep deterministic policy gradient algorithm. Particularly, the HCMAPPO algorithm can be trained with the environment based on the thirteen-turbine wind farm and effectively applied to the larger wind farms. At the same time, there is no significant increase in the fatigue damage of the wind turbine blade from the wake control as the wind farm scale increases. The multivariate HCMAPPO control can realize the collective large-scale wind farm maximum power output.

\end{abstract}


\end{frontmatter}


\section{Introduction}

\subsection{Background and Motivation}
Wind energy is one of the most critical sustainable energy and has become an essential source of global power generation. Recently, the development of wind farms has been growing drastically to harvest more wind power. However, wind farms' power generation efficiency and economic benefits still severely suffer from the high system complexities and uncertain environments. A wind farm power model with a better understanding of wake is crucial to access the variable control strategies.

Some wind farm optimization and control approaches have been proposed to mitigate wake effects \citep{dong2023reinforcement,dong2022wind}. Previous studies maximized wind farm power output by axial induction and yaw angle control. Experimental investigations and analytical models were proposed to explain the wake characteristics and train the control strategies. Extensive wind tunnel tests and field data measurements were carried out\citep{haans2005measurement}. Considering the computational cost, parametric analysis models are proposed for use in the development phase of control strategies. The earliest parametric model was the Jensen model, which defined a top-hat wake model and assumed that the wake velocity defect remained constant within the wake region\citep{jensen1983note}. Since the wake would recover faster to the free flow level near the edge of the zone, the Floris model is proposed\citep{gebraad2016wind}. The Floris model segmented the wake region into three concentric zones to better predict velocity distributions, which calculated the magnitude and direction of the wake centreline as a function of yaw angle and axial induction factor. The Gaussian model changed the wake velocity deficit profile to a Gaussian function, thereby compensating for the shortcomings of the Jensen model, which assumed an unpractical uniform velocity deficit profile\citep{bastankhah2016experimental}. The above parametric models use empirical formulas to calculate wind turbine (WT) power output. Howland et al. corrected this empirical formula by measurement at a wind farm in India\citep{howland2022collective}.

Nowadays, some studies also suggest that the tilt angle has a similar effect on the wake as the yaw angle\citep{wang2020influence}. \citet{padullaparthi2022falcon} proposed a multi-agent deep reinforcement learning (MADRL) based coordinated control for wind farms to maximize energy while minimizing fatigue damage by jointly controlling the pitch and yaw angles of all turbines. In addition, PID control and coordinated model-based predictive control (CMPC) are compared with the proposed MADRL algorithm. A novel composite learning-based controller for each WT was designed to achieve closed-loop yaw tracking, which can guarantee the exponential convergence of tracking errors in the presence of uncertainties of yaw actuators \citep{dong2021intelligent}.

Deep reinforcement learning (DRL) combines reinforcement learning and deep learning algorithms\citep{zhang2022transfer}. There is also work on using DRL for wind farm control, but it is generally less explored and typically uses a single control agent to manage the multiple wind turbines (WTs). A model-free DRL method was proposed to maximize the total power generation of wind farms through the combination of axial induction control and yaw control\citep{xie2021wind}. A knowledge-assisted deep deterministic policy gradient (DDPG) algorithm and three knowledge-assisted learning methods are proposed based on the framework \citep{zhao2020cooperative}. A composite experience replay DDPG approach was applied to optimize wind farms' total power production, considering the strong wake effects among wind turbines and the stochastic features of environments\citep{dong2021composite}.

Recent MARL literature has primarily adopted off-policy learning frameworks, such as multi-agent deep deterministic policy gradient (MADDPG)\citep{lowe2017multi} and value-decomposed Q-learning \citep{sunehag2017value}. \citet{yu2021surprising} demonstrated multi-agent proximal policy optimization (MAPPO) achieves significantly faster run-time and comparable sample complexity to off-policy MARL methods (MADDPG, independent Proximal Policy Optimization\citep{de2020independent}, and QMix\citep{rashid2018qmix}). Learning to communicate effectively among agents has shown to be crucial to strengthen interagent collaboration and ultimately improving the quality of policies learned by MARL. \citet{sheng2022learning} categorized the existing designs for communication topology into five patterns: fully-connected\citep{das2019tarmac}, star\citep{foerster2016learning}, tree\citep{jiang2018learning}, neighboring\citep{jiang2018graph}, hierarchical \citep{sheng2022learning}, and the hierarchical communication topology is the most effective comparing to the other four.

\subsection{Scope and Contributions}

This article proposes a novel large-scale wind farm multivariate control approach for maximizing power output, considering the impact of yaw angles, tilt angles, and axial induction factors of the WTs. The proposed HCMAPPO algorithm addresses multi-dimensional and continuous control problems of the WTs. Furthermore, to further improve cooperation among WTs, the wind farm is divided into multiple aggregators, with hierarchical communication methods employed. Simulation results demonstrate the effectiveness of the multivariate HCMAPPO control in increasing the wind farm power output. More specifically, our contributions to this article are outlined in the following.

First, we present a large-scale wind farm multivariate power model to investigate the impact of WTs wake on power output. The proposed multivariate model is based on actual wind farm data and replaces the traditional empirical model. The model incorporates axial induction control, yaw control, and tilt control.

Second, the HCMAPPO algorithm is proposed to coordinate the multivariate large-scale wind farm continuous controls. The large-scale wind farm is divided into multiple wind turbine aggregators (WTAs), and neighboring WTAs can exchange information through hierarchical communication to maximize the wind farm power output.

Third, simulation results demonstrate that the proposed HCMAPPO algorithm achieves a significant power-increasing performance compared to the existing controls including the MADRL-based controls. Particularly, the HCMAPPO algorithm can be trained with the environment based on the thirteen-turbine wind farm and effectively applied to the larger wind farms. 

Finally, the fatigue damage of the WT blade under the proposed control algorithm almost stays the same as wind farms scale up.
\section{Problem Formulation}
The collective large-scale wind farm multivariate control problem is introduced in this section. The WTs in the wind farm are divided into $M$ WTAs according to their spatial positions, and each WTA contains $N$ WTs. Then the steady-state power output of the $n$th WT in the $m$th WTA $WT_n^m$ (denoted by $P_n^m$). The coordinates of the $WT_n^m$ in the wind farm are defined as $[X_n^m, Y_n^m]$. Therefore, considering the influence of yaw angle $\gamma_n^m$ and tilt angle $\beta_n^m$ and axial induction factor $\alpha_n$ of $WT_n^m$, the steady state power output ${P}_{n}^m$ of the $WT_n^m$ in the wake region is expressed as,
\begin{align}
{{P}_{n}^m}=\frac{1}{2}{{C}_{p}}\left( \alpha_n^m \right){{C}_{\gamma }}\left( \gamma_n^m \right){{C}_{\beta }}(\beta_n^m )\rho {A_n^m}{u_n^m}^{3}, \label{pwt}
\end{align}
where $\rho$ is the air density, and $A_n^m$ is the rotor area of $WT_n^m$, $A_n^m=\frac{1}{4}\pi {{D_n^m}^{2}}$. ${D_n^m}$ is the diameter of the WT rotor plane of $WT_n^m$. $u_n^m$ is the wind speed in front of $WT_n^m$. The $C_{\gamma }(\gamma_n^m)$ (refer to Eq.~(\ref{Cbeta}) in \ref{power}) is obtained by fitting the statistic experimental data from \citep{howland2022collective}. The tilt coefficient ${C_{\beta }(\beta_n^m)}$ is obtained by fitting the experimental data from \citep{wang2020influence}. The empirical formula of yaw angle on power output is replaced by the ${C}_{\gamma }(\gamma_n^m)$ (refer to Eq.~(\ref{Cgamma}) in \ref{power})). The power coefficient ${C}_{p}(\alpha_n^m)$ is decided by the axial induction factor $\alpha_n^m$. ${C}_{p}(\alpha_n^m)$ satisfies:
\begin{align}
{{C}_{p}}\left( {{\alpha}_{n}^m} \right)\text{=4}{\alpha_{n}^m}{{\text{(1-}{\alpha_{n}^m}\text{)}}^{2}}.
\end{align}
The steady state power model details are presented in~\ref{power}.

\begin{figure}[htb]
\centering
\includegraphics[width=0.9\textwidth]{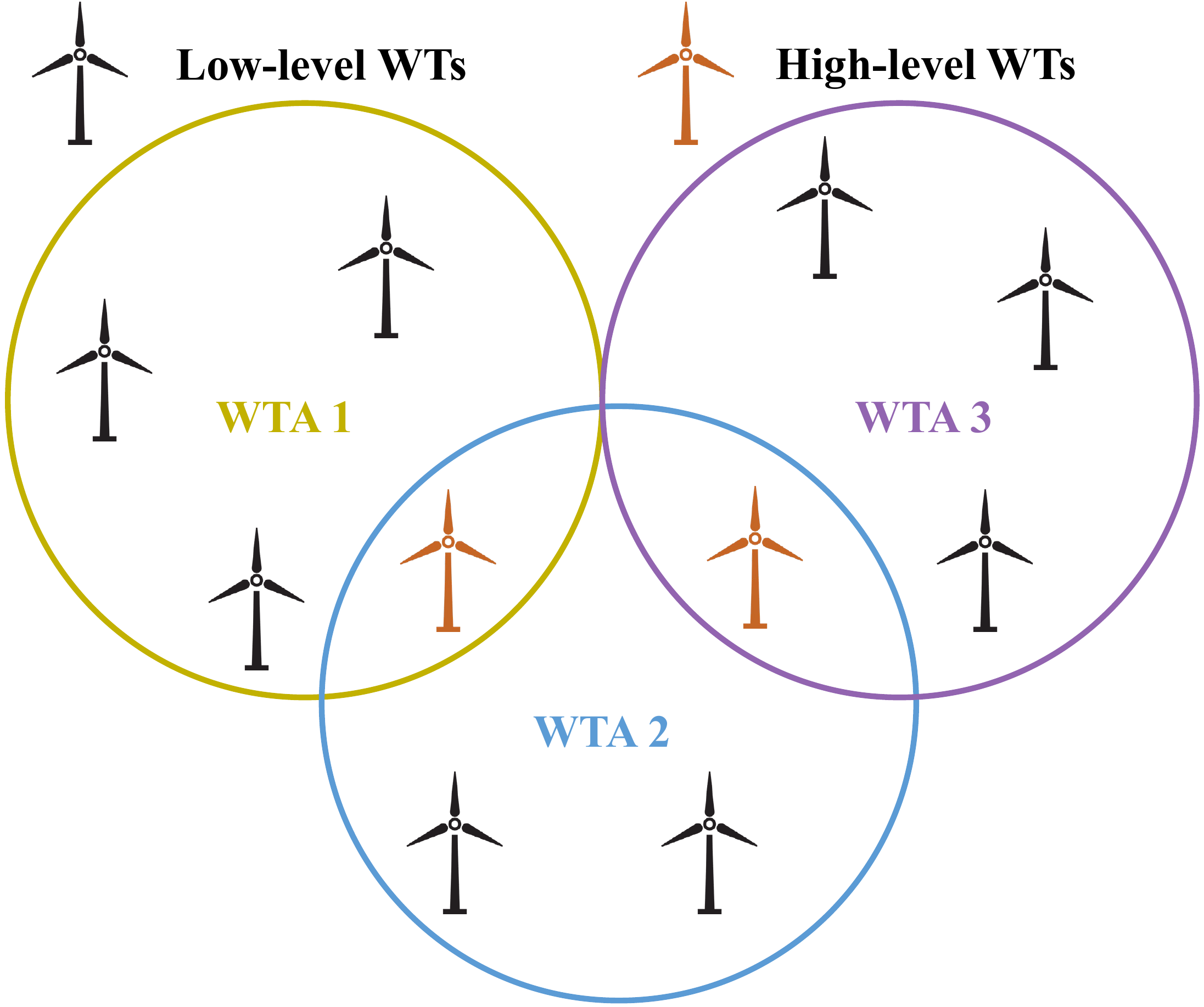}
\caption{The illustrative structure of WTAs in an exemplary ten-turbine wind farm.}
\label{fig:wta}
\end{figure}

In controlling large-scale wind farm power output, the wind farm is divided into the WTAs to realize collective wind farm control. The WTA framework is shown in Figure~\ref{fig:wta}. Except for the $M$th WTA, the $N$th WT in each WTA is considered a high-level WT. Alternatively stated, except for the first WTA, the first WT in each WTA is a high-level WT. Each high-level WT is located between two neighboring WTAs and can transfer information between the two neighboring WTAs.
Considering the coupling relationship between WTAs in the wind farm, the wind farm power output $P_{WF}$ is
\begin{align}
P_{WF}=\sum\limits_{m=1}^{M}P_{EVA}^m-\sum\limits_{m=2}^{M}{P}_{n=1}^{m}\label{pwf},
\end{align}
where $P_{EVA}^m$ is power output of the $m$th EVA, $P_{EVA}^m =\sum\limits_{n=1}^{N}{P_n^m}\label{peva}$
and ${P}_{n=1}^{m}$ is the power output of the first $WT$ in the $m$th EVA. The first WT in the $m$th WTA, $WT_{n=1}^m$ and the $N$th WT in the $m-1$th WTA, $WT_{n=N}^{m-1}$ are the same high-level WT. The shared WTs between two WTAs are the high-level blue-colored WTs shown in Figure~\ref{fig:wta}, utilized for communication among neighboring WTAs. According to the wind farm power model presented in (\ref{pwt})-(\ref{pwf}), the optimization problems formulated for the wind farm power control in the form of WTAs can be described as,
\begin{equation}
\max\limits_{{\gamma }_{n}^m,{\beta }_{n}^m,{\alpha}_{n}^m}{P_{WF}},\label{goal}
\end{equation}
subject to:
\begin{align}
&-45^{\circ}\le {{\gamma }_{n}^m}\le 45^{\circ},\\
&-15^{\circ}\le {{\beta }_{n}^m}\le 15^{\circ},\\
&0\le {\alpha_{n}^m}\le \frac{1}{3}.\label{cons}
\end{align}
In the multivariate wind farm control, all three factors, WT's yaw angle, tilt angle, and axial induction factor, are continuously adjusted to mitigate the wake effects on downstream WTs and maximize the wind farm power output.

\section{HCMAPPO-Based Wind Farm Control}
It is a challenging task to rapidly solve the wind farm multivariate optimization problem to maximize the power output using the classic methods\citep{padullaparthi2022falcon} due to the following reasons
\begin{enumerate}
\item The wind farm environment is nonlinear and the power output optimization problem is a non-convex optimization problem.
\item Each WT are affected by the wake effects of other WTs. As indicated by (\ref{pwt})-(\ref{cons}), the problem formulated for the wind farm power is coupled with the WTAs in $M$.
\item Considering the sensitive delay requirements of wind farm power output control, each formulated problem has to be solved rapidly.
\end{enumerate}
The communication-based MADRL approach is leveraged here. Each WTA is an agent to learn the power output maximization. Specifically, we re-model the wind farm multivariate optimization in terms of $M$ WTAs as a multi-agent extension of partially observable Markov decision process processes (POMDPs). The HCMAPPO algorithm is developed to solve the POMDPs.

\subsection{Problem Transformation}
We transform the multivariate control problems into POMDPs for the $M$ WTA agents. Define the POMDPs for $M$ agents as a set of states $\mathcal{S}$, a set of observations $\mathcal{O}=\left\{o^1, \ldots, o^m, \ldots, o^M\right\}$, and a set of actions $\mathcal{A}=\left\{{a}^1, \ldots, a^m, \ldots, a^M\right\}$. The state set $\mathcal{S}$ describes the wind farm. $\mathcal{O}^m$ are observation spaces for the $m$th WTA agent $(m \in M)$, and the observation of each agent at timeslot $t$ is a part of the current state, $s(t) \in \mathcal{S}$. $\mathcal{A}^m(m \in M)$ are action spaces for the $m$th WTA. For each given state $s \in \mathcal{S}$, the $m$th WTA agent uses the policies $\pi: \mathcal{S} \mapsto a^m$, to choose an action from their action spaces according to their observations corresponding to $o^m$, respectively.

We detail the communication-based MADRL formulation of the examined wind farm control problem, the critical elements outlined in the following.

\paragraph{Environment} The environment consists of the WTs, all of which interact in the form of WTAs. Moreover, neighboring WTAs can communicate with each other and transmit information.

\paragraph{Agent and communication} The WTA constitutes the agent, which gradually learns how to improve its decisions by utilizing experiences from repeated interactions with the wind farm environment.

\begin{figure*}[htb]
\centering
\includegraphics[width=1.0\textwidth]{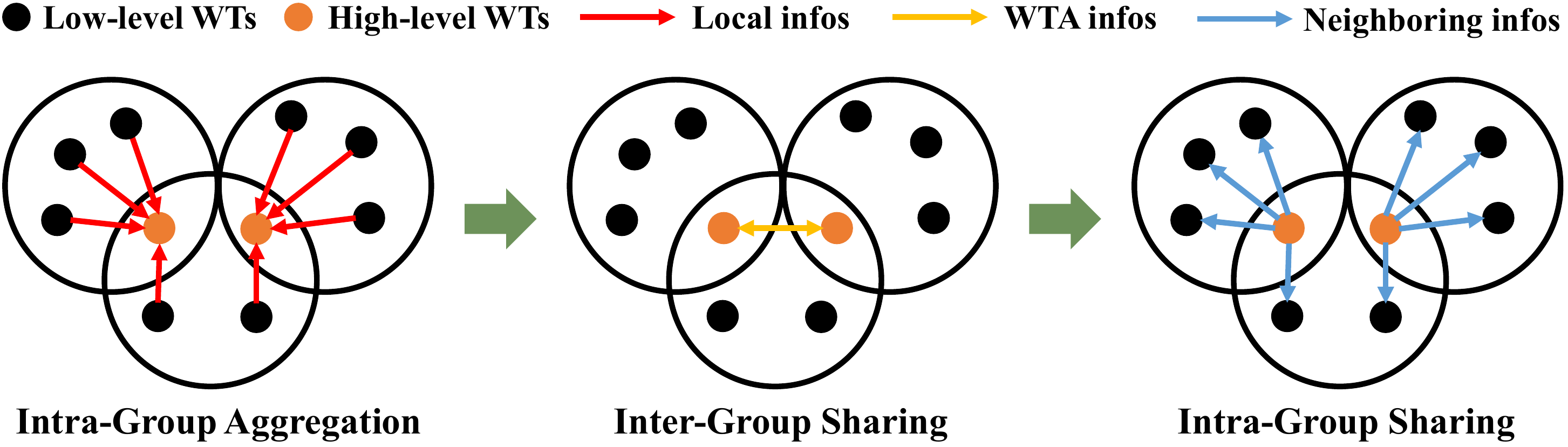}
\caption{WTAs' hierarchical communication in the ten-turbine wind farm. "Local infos" represents the wind speed and direction transmitted from lower-level WTs within a WTA to high-level WTs. "WTA infos" denotes the observations made by the WTA agent, specifically the agent's local observation. "Neighboring infos" indicates the information exchanged among neighboring WTAs and subsequently transmitted to the lower-level WTs within the WTAs.}
\label{fig:comm}
\end{figure*}

To improve both efficient message accessibility and adequate message comprehension for large-scale WTAs, the hierarchical communication topology is adopted in wind farm control. As shown in Figure~\ref{fig:comm}, the overall hierarchical communication sub-module consists of three steps.
\begin{itemize}
\item Step 1) Intra-group aggregation. The low-level WTs embed their local information in each WTA and send it to the high-level WTs. The high-level WTs aggregate the information from all associated low-level WTs and obtain the WTA perception.

\item Step 2) Inter-group sharing. The high-level WTs communicate with the neighboring high-level WTs with WTA perception. It is unnecessary to aggregate all received other WTs messages further to obtain the global perception, reducing the pressure on communication.

\item Step 3) Intra-group sharing. Each high-level WT communicates all its features with the associated low-level WTs, while the low-level WTs aggregate the received information from high-level WTs. The embedding features of both high-level and low-level WTs are then updated.
\end{itemize}
The hierarchical communication topology does not require global communication and will not cause the interference of redundant information. Thus, the high efficiency of communication can be guaranteed.

\paragraph{Observation} Due to the ability of WTA agents to exchange information with neighboring agents, the observations of the $m$th WTA agent at timeslot $t$ $o^m(t)\in O^m$ can be described as
\begin{align}
&{{o}^{m}}(t)=\left\{ {{x}_{1}^m}(t),{{x}_{2}^m}(t),\ldots ,{{x}_{N}^m}(t),{{y}_{1}^m}(t),{{y}_{2}^m}(t),\ldots , {{y}_{N}^m}(t),\right.\nonumber\\
&\qquad \qquad \left.u_{1}^m(t),u_{2}^m(t),\ldots ,u_{N}^m(t), \right.\nonumber\\
&\qquad \qquad \left.wd_{1}^m(t),wd_{2}^m(t),\ldots ,wd_{N}^m(t) , {\zeta}^{m}(t) \right\},\label{oj}\\
&{\zeta}^{m}(t)=\left\{\zeta_{1}^m(t),\zeta_{2}^m(t),\ldots ,\zeta_{N}^m(t) \right\},
\end{align}
where ${{x}_{1}^m}(t),{{x}_{2}^m}(t),\ldots ,{{x}_{N}^m}(t)$ and ${{y}_{1}^m}(t),{{y}_{2}^m}(t),\ldots , {{y}_{N}^m}(t)$ represent $x$ and $y$ coordinates of $N$ WTs in $m$th WTA at timeslot $t$, respectively. $u_{1}^m(t),u_{2}^m(t),\ldots ,u_{N}^m(t)$ and $wd_{1}^m(t),wd_{2}^m(t),\ldots ,wd_{N}^m(t)$ indicate the wind speed and direction of $N$ WTs in $m$th WTA at timeslot $t$, respectively. ${\zeta}^{m}(t)$ are the messages received by the $m$th WTA at timeslot $t$, always the other WTA agent's wind speed and direction.

\paragraph{State}

The state space $\mathcal{S}$ encapsulates $M$ WTAs' observation. The state at timeslot $t$ is a vector $\mathcal{S}$. The state of the $m$th WTA at timeslot $t$, i.e., $s(t)\in \mathcal{S}$, can be described as

\begin{align}
s(t)=\left\{o^1(t), \ldots, o^m(t), \ldots, o^M(t)\right\}.
\end{align}

\paragraph{Action}

According to the current policy $\pi$ and the corresponding observation, each WTA agent chooses an action from its action space. The actions of the $m$th WTA agent at timeslot t, i.e., $a^m(t)\in \mathcal{A}^m,$ $(m\in M)$, can be described as

\begin{align}
&a^m(t)=\left\{ {\alpha_{1}^m}(t),{\alpha_{2}^m}(t),\ldots ,{\alpha_{N}^m}(t),{\beta_{1}^m}(t),{\beta_{2}^m}(t),\ldots , {\beta_{N}^m}(t),\right.\nonumber\\
&\qquad \quad \left.\gamma_{1}^m(t),\gamma_{2}^m(t),\ldots ,\gamma_{N}^m(t), b^m(t).\right\},\label{aj}\\
&{b}^{m}(t)=\left\{ b_{1}^m(t),b_{2}^m(t),\ldots ,b_{N}^m(t) \right\}.
\end{align}

The action $a^m(t)$ represents the control variables and messages of the $m$th WTA at the timeslot $t$. $a^m(t)$ includes the yaw angles, tilt angles, and axial induction factors of WTs in the $m$th WTA. ${b}^{m}(t)$ is the local messages sent by the $m$th WTA at timeslot $t$, which is always the WTs' wind speed and direction in the $m$th WTA.

\paragraph{Reward}
The reward is the optimization goal for the WTA agents. To avoid sparse rewards during the agent learning process, the functions below highlight the rewards and penalties for every action undertaken by the agent while scheduling WTA.

\begin{align}
&r^m=P_{WF},\label{rr}
\end{align}
where $r^m$ is the reward of the $m$th WTA agent. The reward function is based on (\ref{pwt})-(\ref{cons}), and the goal is to maximize the reward.

\subsection{MAPPO}

MAPPO trains two separate neural networks: an actor network with parameters $\theta$ and a critic network with parameters $\phi$ \citep{yu2021surprising}. Specifically, the critic network, denoted as $V_{\phi}$, performs the following mapping: $\mathcal{S}\mapsto \mathcal{R}$. The global state can be agent-specific or agent-agnostic.

The actor network, denoted as $\pi_{\theta}$, maps agent observations $o^m$ to a categorical distribution over actions to the mean and standard deviation vectors of a Multivariate Gaussian Distribution, from which an action is sampled, in continuous action spaces. The actor and critic networks are trained to maximize the loss functions $L(\theta)$ and $L(\phi)$, respectively. Regarding further details about the MAPPO algorithm, please refer to \citep{yu2021surprising}.

\subsection{HCMAPPO-Based Solution}

\begin{figure*}[htb]
\centering
\includegraphics[width=1.0\textwidth]{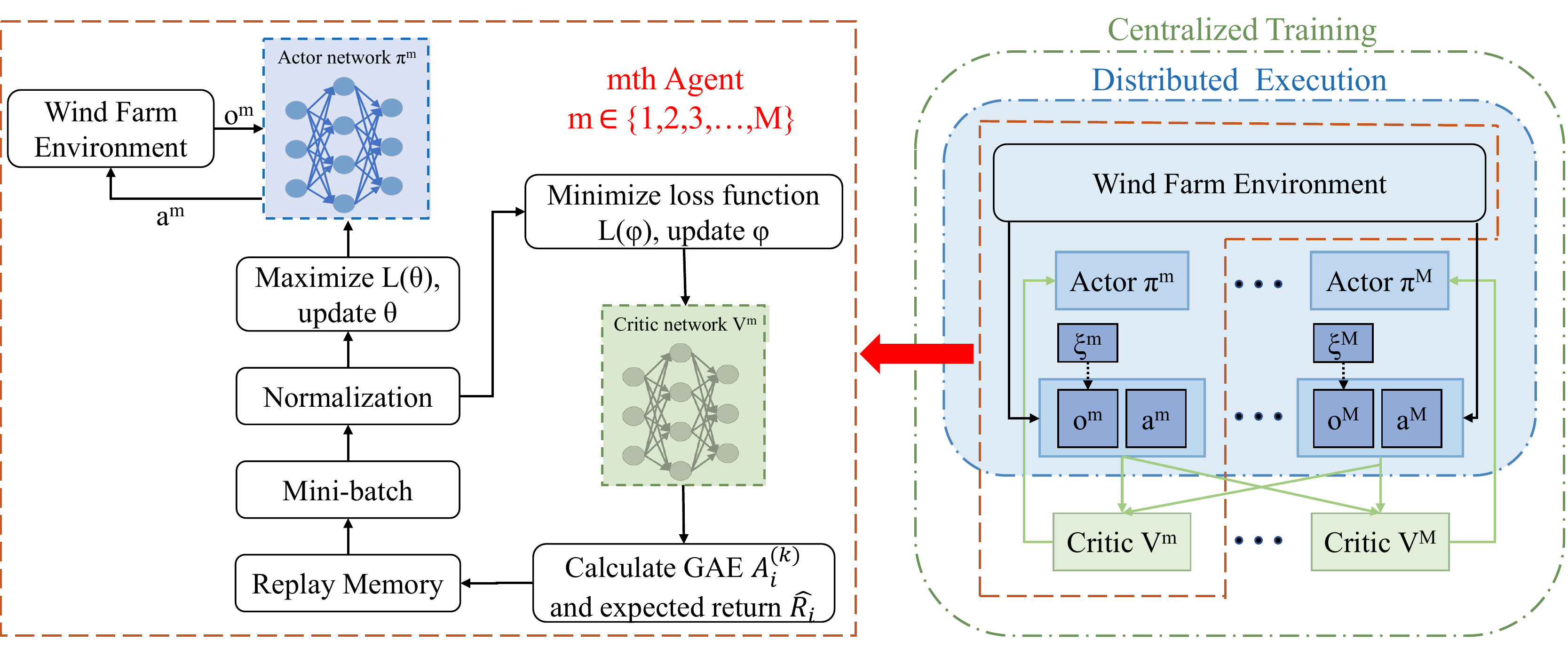}
\caption{The centralized training and distributed execution framework of the HCMAPPO algorithm in the wind farm.}
\label{fig:mappo}
\end{figure*}

The centralized training and distributed execution framework of the HCMAPPO algorithm in the wind farm is illustrated in Figure~\ref{fig:mappo}. The wind farm environment and $M$ collaborative WTA agents interact to maximize the wind farm power output in the HCMAPPO algorithm.

From the actor and the critic compositions, centralized training and decentralized execution can be adopted directly in the MAPPO algorithm\citep{foerster2016learning}. Next, we take the $m$th WTA agent as an example to explain how to centrally train the HCMAPPO model and execute the learned model in a decentralized way.

In the centralized training stage, in addition to the local observation $o^{m}$, communication messages are also available to the $m$th agent. When updating the parameters of the actor and the critic network according to the inputted mini-batch of transitions, the actor network chooses an action $a^{m}$ according to the local observation $o^{m}$. And the chosen action $a^{m}={\mu}^{m}\left(o^{m}\right)$ and $s^{m}$ are valued by the critic network. The power of each WT is co-determined by the actions of the $M$ WTA agents, using $s^{m}$, which includes the messages about other agents' actions to learn the state-action value function $V$ would ease training. Moreover, with the communication messages, each agent allows learning its state-action value function separately. Also, as aware of all other agents' actions, the environment is stationary to each agent during the centralized training stage. Thus, the dynamic environment caused by other agents' actions in other MADRL algorithms has been addressed through centralized training and distributed execution. During the execution stage, as the actor network requires only local observation, including the neighboring agents' communication messages, each agent can obtain its action.

Considering the common objective of the formulated optimization problems, the $M$ agents should cooperatively maximize the wind farm power output. Each agent collaborates to achieve a cooperative Markov game by setting an identical immediate reward as (\ref{rr}).

\section{Case Study}

\subsection{Test System and Implementation}

The effective wind farm operation is a challenging and complex task due to its distributed and dynamic nature. Coordinated control of individual WTs in the large-scale wind farm using the HCMAPPO algorithm can significantly improve the efficiency and reliability of large-scale wind farm operations. The hyperparameters of the HCMAPPO algorithm are listed in Table~\ref{tab:hmappo}. 
\begin{table}[h]
 \centering
 \caption{HCMAPPO training hyperparameters}
 \label{tab:hmappo}
 \begin{tabular}{cc}
  \toprule
\makecell[c]{Hyperparameter}   &\makecell[c]{Value}   \\ 
  \midrule
   Number of episodes     & $4*10^4$ \\
   Length of episode  & 25   \\
   Discount factor $\gamma$    & 0.99   \\
   Learning rate for actor    & $5*10^{-4}$   \\
   Learning rate for critic    & $5*10^{-4}$   \\
   Gain of last action layer    & $0.01$   \\
   Entropy term coefficient   & $0.01$   \\
 Actor update steps       &15\\
 Critic update steps       &15\\
Dimension of states       &20\\
 Dimension of action       &24\\
Epsilon $\epsilon$    & 0.2\\
 RMSprop optimizer epsilon   & $1*10^{-5}$\\
 GAE lambda parameter   & $0.95$\\
  \bottomrule
 \end{tabular}
\end{table} 

\begin{figure}[h]
\centering
\includegraphics[width=0.9\textwidth]{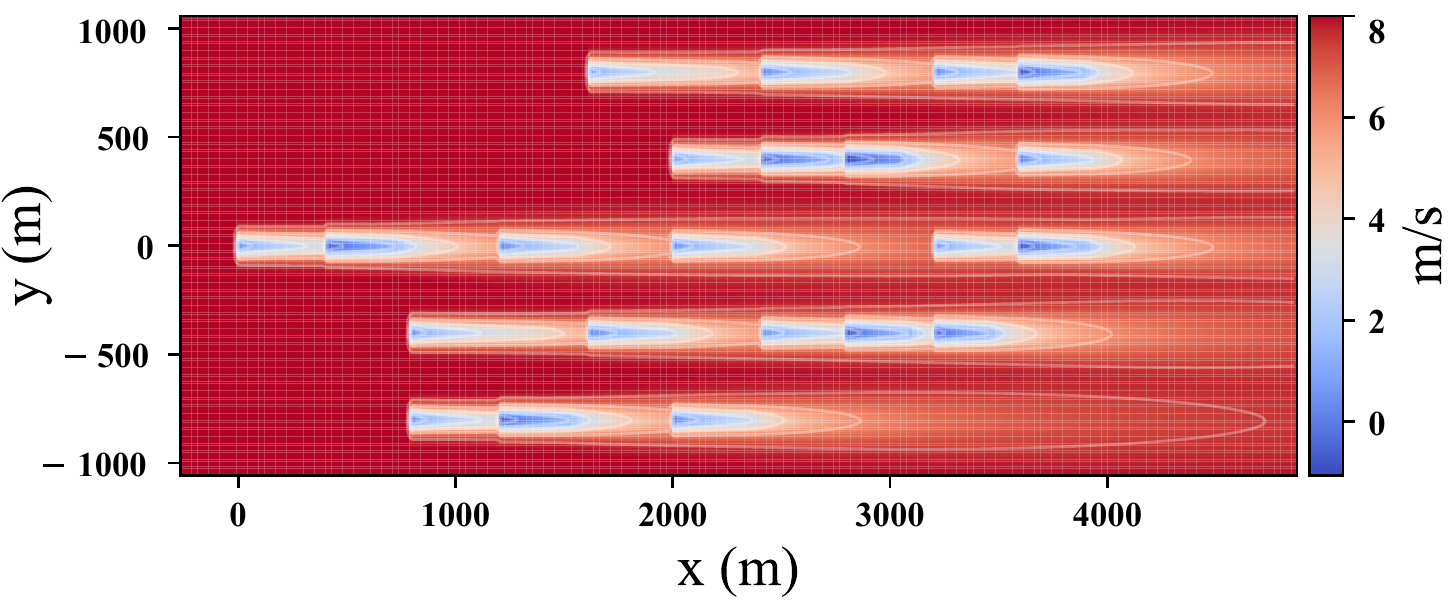}
\caption{The flow field in the twenty-two-turbine wind farm.}
\label{fig:wf}
\end{figure}

The study considers real-world wind farms with 13, 16, 19, and 22 WTs (layout presented in Table~\ref{tab:dlwf} in \ref{layout}), where each WT has a 5 MW rating. The free-stream velocity $u_{\infty}$ is 8 m/s, and the direction is positive along the X-axis. The flow field in the twenty-two-turbine wind farm is shown in Figure~\ref{fig:wf}, and the influence of the wake effects can be observed. The change of color implies the change in wind speed at different locations, and the magnitude of the minimum and maximum wind speeds are $0 m/s$ (blue color just behind the turbines) and $8 m/s$ (red color). The white lines are the wind speed contours. The wind speed behind each WT drops sharply and gradually returns to free-stream velocity. When WTs are aligned, the wind speed of the downstream WTs is lower than that of the upstream WTs. The region shaded with a deep blue hue in Figure~\ref{fig:wf} illustrates the influence imparted by the wake impact stemming from the upstream WTs onto the downstream WTs.

\subsection{Large-scale wind farm control}

\begin{figure}[h]
\centering
\includegraphics[width=0.49\textwidth]{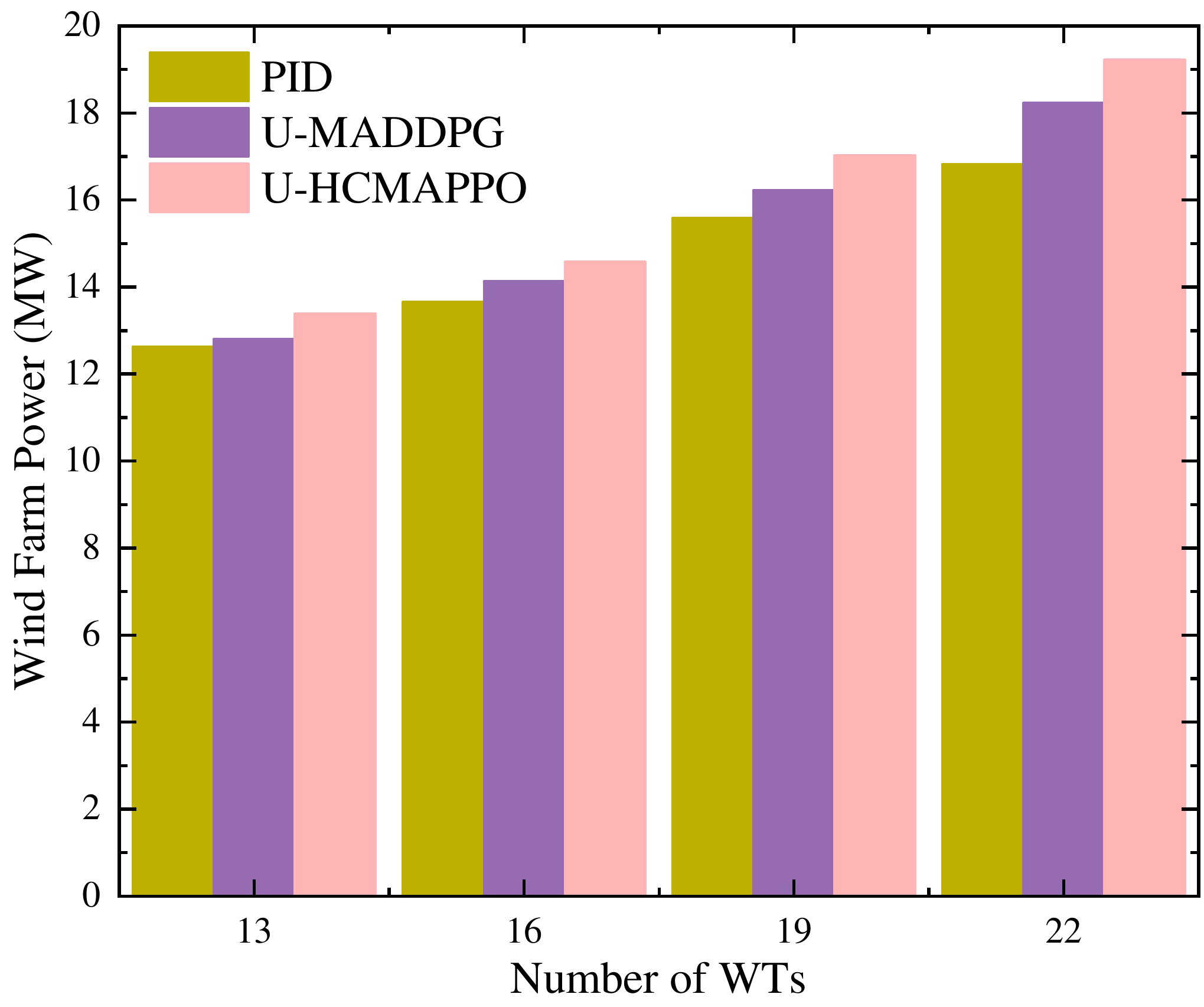}
\caption{The power outputs of the four different wind farms based on the three univariate controls.}
\label{fig:wtpall2}
\end{figure}

\begin{figure}[h]
\centering
\includegraphics[width=0.49\textwidth]{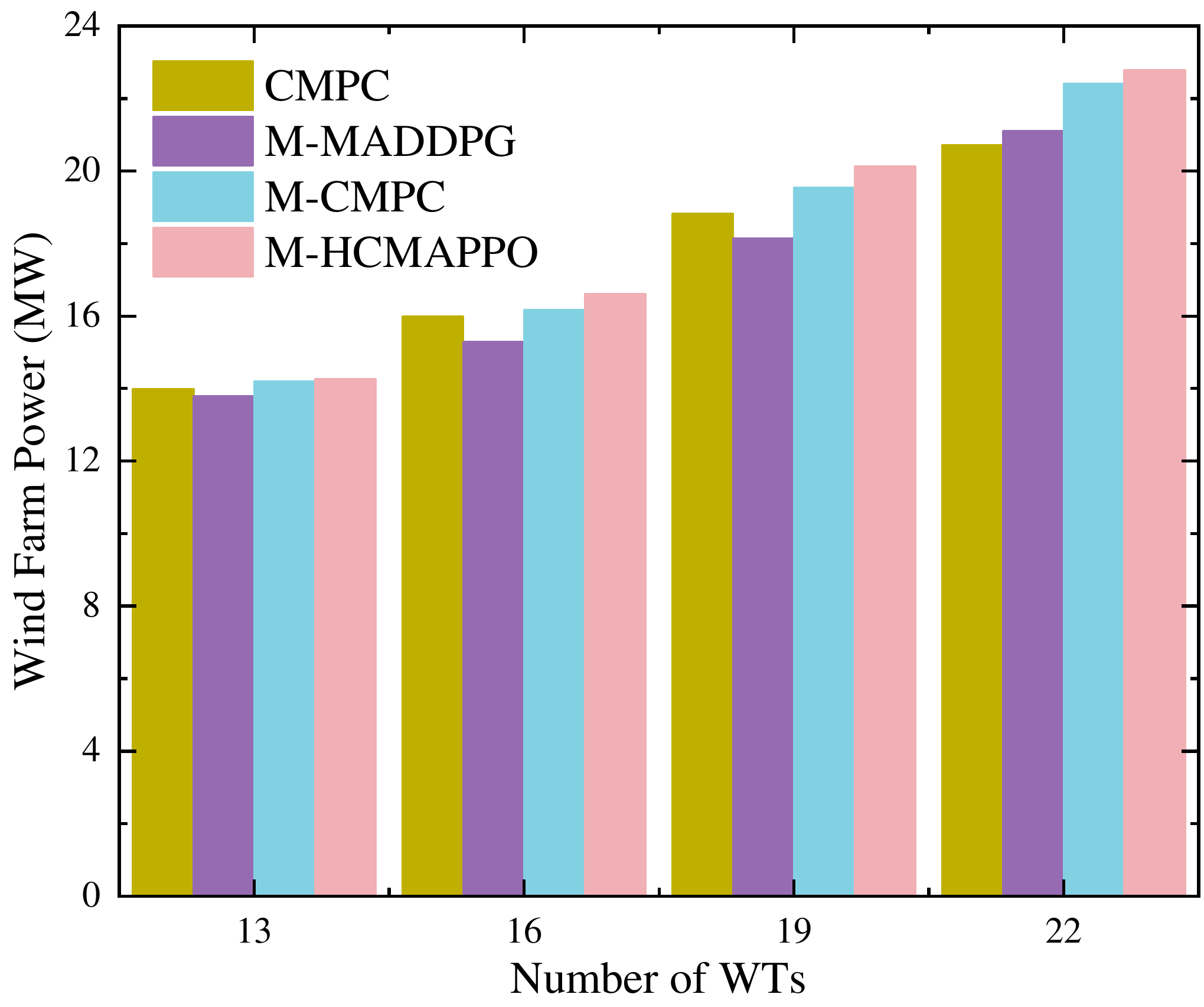}
\caption{The power outputs of the four different wind farms for the M-HCMAPPO, CMPC, M-MADDPG, and M-CMPC four multivariate controls.}
\label{fig:wtpall3}
\end{figure}

The univariate control typically has the yaw angle control variable, while the multivariate control has three control variables: axial induction factors, yaw angles, and tilt angles.
Based on the HCMAPPO algorithm, the univariate wind farm control (U-HCMAPPO) and the multivariate wind farm control (M-HCMAPPO) are tested for the four wind farms with 13, 16, 19, and 22 WTs. The CMPC algorithm with the bivariate control of WTs' yaw and pitch angles is used as the optimal benchmarking control~\cite{padullaparthi2022falcon}.
Correspondingly, as the univariate control, the U-HCMAPPO is compared with the PID algorithm with the univariate yaw control (PID)~\cite{padullaparthi2022falcon}, the MADDPG algorithm~\cite{lowe2017multi} with univariate control (U-MADDPG) as shown in Figure~\ref{fig:wtpall2}.
As the multivariate control, the M-HCMAPPO control is compared with the MADDPG algorithm with the multivariate control (M-MADDPG) and the CMPC algorithm with the multivariate controls (M-CMPC). The power outputs of different wind farms are shown in Figures ~\ref{fig:wtpall2} and ~\ref{fig:wtpall3} for the univariate and multivariate controls, respectively.
The U-HCMAPPO control can optimally and effectively coordinate the WTs in the wind farms and generate more power output compared to the PID control and U-MADDPG control, as shown in Figure~\ref{fig:wtpall2}.
As for the multivariate control, the M-HCMAPPO control generates more power output than the benchmarking bivariate CMPC control\cite{padullaparthi2022falcon}, the M-MADDPG control, and the M-CMPC control in the four wind farms with 13, 16, 19 and 22 WTs as shown in Figure~\ref{fig:wtpall3} because the HCMAPPO algorithm's hierarchical communication promotes coordination among WTA agents.

Compared with Figure~\ref{fig:wtpall2}, the M-HCMAPPO control generates more power than the U-HCMAPPO controls due to the three control variables in the M-HCMAPPO control. The M-HCMAPPO control outperforms the M-CMPC due to its continuous nature. Since the M-CMPC employs a discrete variable traversal mechanism, it results in a relatively long solution time, which does not suit the actual operational needs. The proposed M-HCMAPPO control operates continuously and provides a more optimal control strategy than CMPC and M-CMPC without sacrificing computation time. As a result, the collective wind farm multivariate control based on the HCMAPPO algorithm is the most effective control strategy in wind farm power production.

\begin{figure}[h]
\centering
\includegraphics[width=0.5\textwidth]{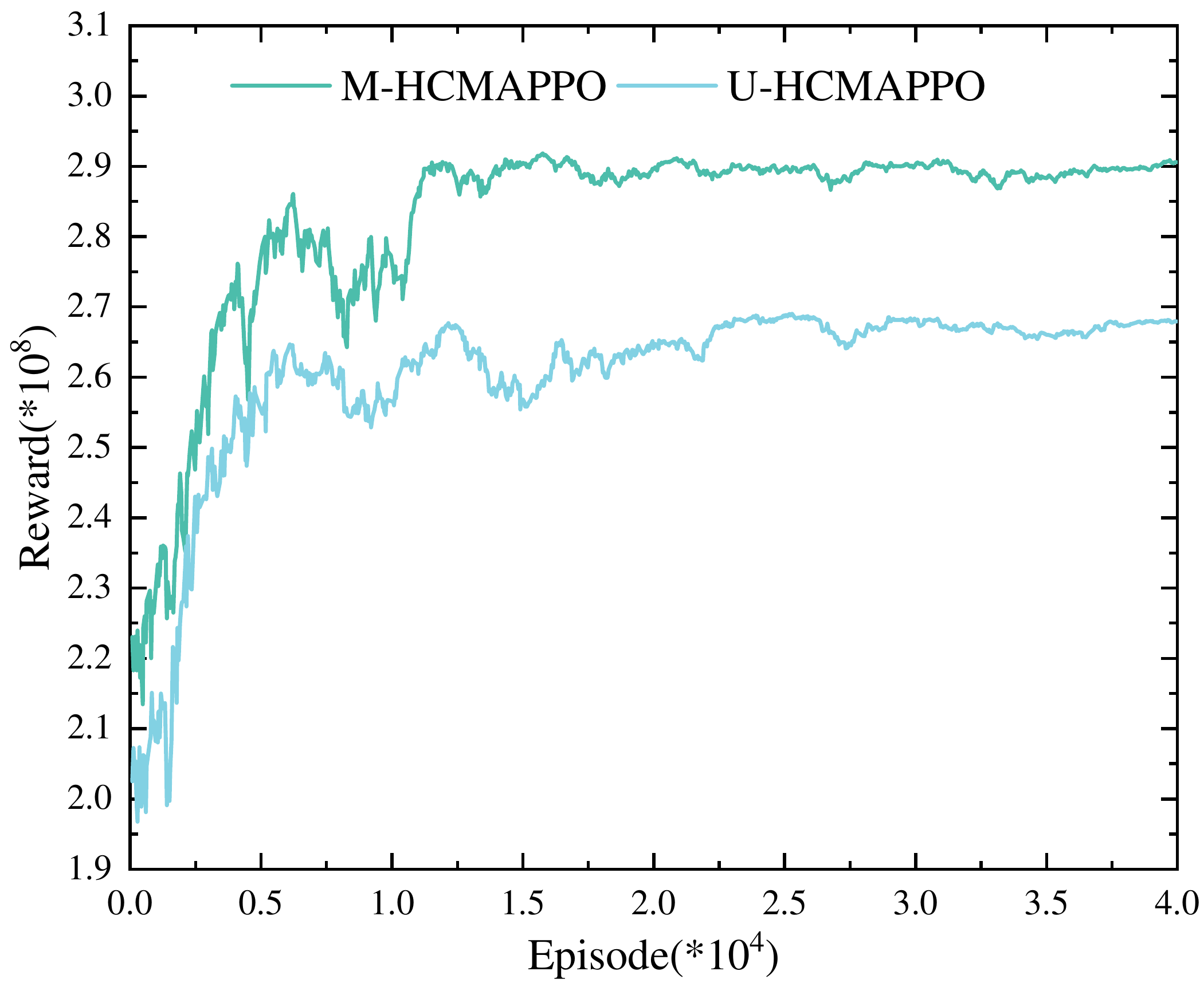}
\caption{The rewards of the HCMAPPO algorithm in the thirteen-turbine wind farm.}
\label{fig:rewardmappo}
\end{figure}

\begin{figure}[h]
\centering
\includegraphics[width=0.9\textwidth]{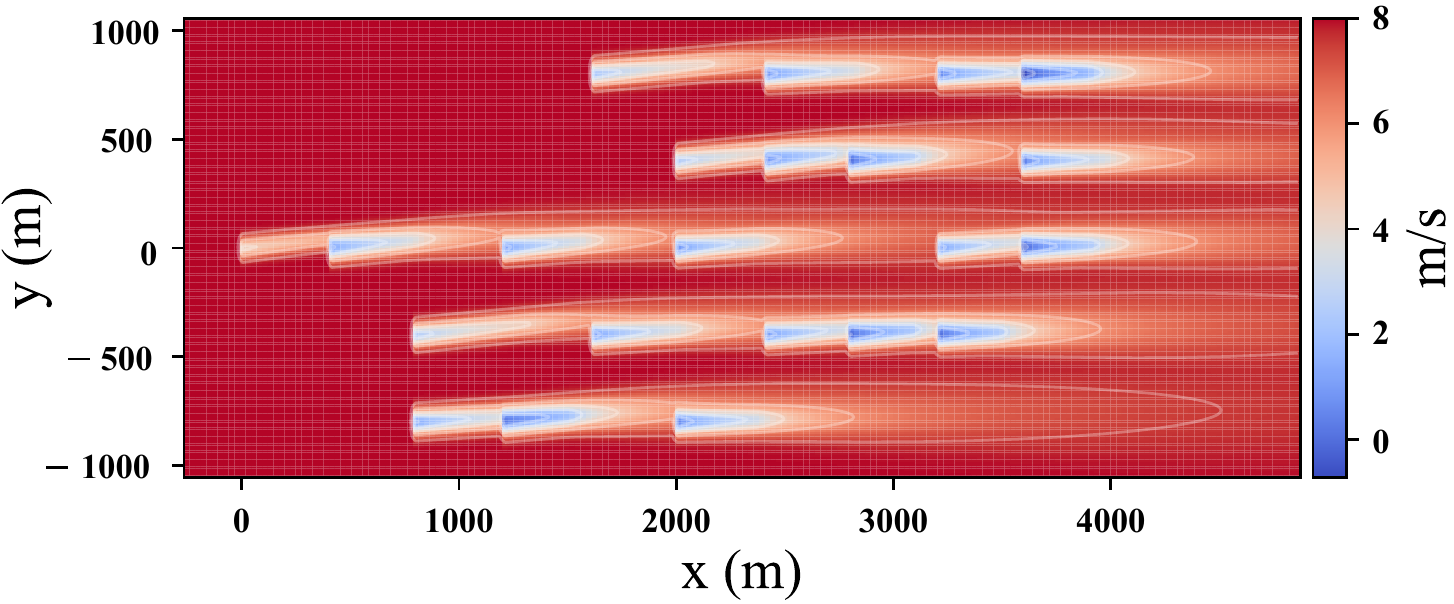}
\caption{The flow field in the twenty-two-turbine wind farm under M-HCMAPPO control.}
\label{fig:yrLHC22}
\end{figure}

The rewards of the HCMAPPO algorithm on the thirteen-turbine wind farm are shown in Figure~\ref{fig:rewardmappo}.
The HCMAPPO algorithm has the trained parameters in Table~\ref{tab:hmappo}. The M-HCMAPPO control clearly gains more rewards in the model training than the U-HCMAPPO control, which means the M-HCMAPPO control can control wind farms more optimally. Although the HCMAPPO algorithm is trained on the thirteen-turbine wind farm, the HCMAPPO algorithm can be directly transferred to the wind farms with 16, 19 and 22 WTs. Essentially, the HCMAPPO algorithm is potentially trained on small data samples and transferred to large-scale wind farms. The M-HCMAPPO demonstrates an attractive advantage in terms of cost-effective computation to train the M-HCMAPPO control that can be effectively applied to large-scale wind farms given the strong generalization ability for wind farms. Taking the twenty-two-turbine wind farm as an example, Figure~\ref{fig:yrLHC22} presents the flow field in the twenty-two-turbine wind farm under the M-HCMAPPO control.

The M-HCMAPPO control can successfully mitigate the wake effects by controlling the three control variables of WTs' yaw angles, tilt angles, and axial induction factors to increase wind speed at the location of downstream WTs.
The power output produced by the M-HCMAPPO control reaches an impressive 22.79 MW, providing compelling evidence of the efficacy of the M-HCMAPPO control.
The M-HCMAPPO control produces 35.41\% more wind farm power output than the univariate PID control and 9.99\% more wind farm power output than the CMPC control. Additionally, the M-HCMAPPO control can produce 7.96\% more power output than the M-MADDPG control and 1.70\% more than M-CMPC.
In summary, the M-HCMAPPO control can generate more wind farm power output than the four multivariate controls for the wind farms with 13, 16, 19 and 22 WTs.

\subsection{Average damage equivalent load}
Based on the fatigue damage of WT blade~\cite{meng2023study}, the average damage equivalent load (DEL) of single WT per unit of power output $\overline{D}$ is defined as,
\begin{align}
&\overline{D}=\sum\limits_{n=1}^{N}{\sqrt{{DEL_{y,n}}^2+{DEL_{x,n}}^2}}/P_{WF},\\
&DEL_{y,n}=0.01\gamma^2_n-0.59\gamma_n+585.6,\label{DELy}\\
&DEL_{x,n}=-0.07\gamma_n^2+3.99\gamma_n+76.69,\label{DELx}
\end{align}
where, the $DEL_{y,n}$ and $DEL_{x,n}$ are the DEL along the in-plane and out-of-plane directions of the $n$th WT.
The average damage equivalent load can be used to study the wake effects quantitatively for wind farms.

\begin{figure}[h]
\centering
\includegraphics[width=0.49\textwidth]{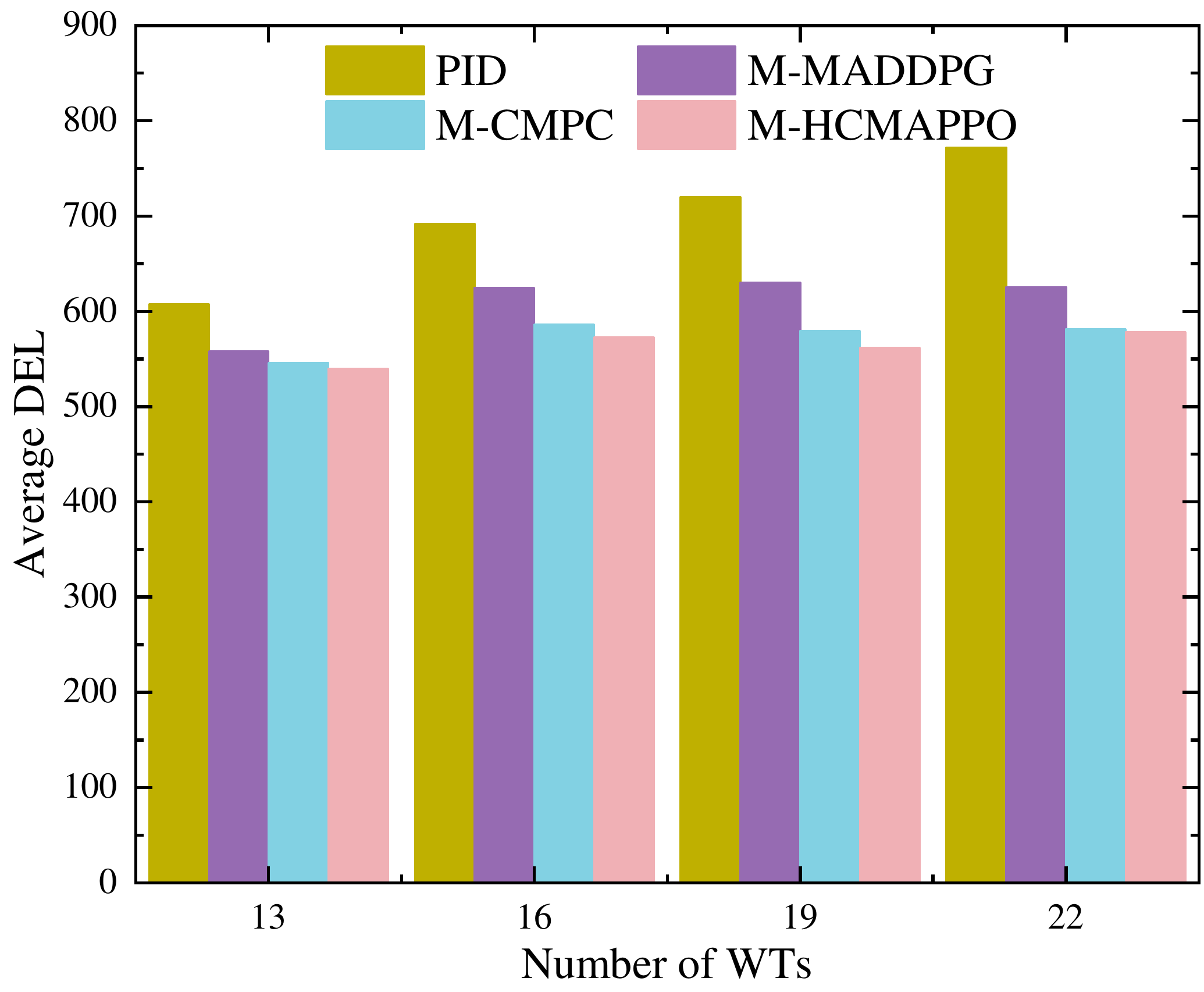}
\caption{The average damage equivalent load across the four different wind farms for the M-HCMAPPO, PID, M-MADDPG, and M-CMPC four controls.}
\label{fig:DEL}
\end{figure}

Figure~\ref{fig:DEL} shows the $\overline{D}$ for the PID, M-MADDPG, M-CMPC, and M-HCMAPPO four controls for the four different wind farms. As the wind farm scales up from 13 WTs to 22 WTs, the $\overline{D}$ for all the four control algorithms increases. However, the M-HCMAPPO control shows the smallest increase of $\overline{D}$ as the wind farm scales up compared to those of PID control, M-MADDPG control, and M-CMPC.
\section{Conclusion}
The HCMAPPO algorithm is proposed in the communication-based MADRL domain for the large-scale WT continuous multivariate coordination in the wind farm control. The hierarchical communication between WTAs can schedule large-scale WTs efficiently. Then, taking the maximum wind farm power output as the main objective, the wake effects between WTs are considered. Finally, the HCMAPPO algorithm was adapted to optimize the WTs' axial induction control, yaw control, and tilt control in continuous actions. Compared with traditional control algorithms (e.g., PID and CMPC) and traditional MADRL algorithms (e.g., MADDPG), the collective wind farm multivariate control based on the HCMAPPO algorithm is superior. Therefore, the HCMAPPO algorithm is suitable for large-scale wind farm control and independent of prediction information.

In future work, the realistic wind farm data can be used to make the large-scale wind farm multivariate power model more accurate and achieve the transient controls.

\section*{Funding}
This work was supported in part by the National Natural Science Foundation of China (Grant No.21773182 (B030103)) and the HPC Platform, Xi'an Jiaotong University.

\section*{CRediT authorship contribution statement}

\textbf{Yubao Zhang:} Validation; Roles/Writing – original draft; Formal analysis; Data curation. 
\textbf{Xin Chen:} Conceptualization; Methodology; Writing – review \& editing; Supervision.
\textbf{Sumei Gong:} Software; Writing – review \& editing.
\textbf{Haojie Chen:} Supervision.

\section*{Declaration of competing interest}

The authors declare that the research was conducted in the absence of any commercial or financial relationships that could be construed as a potential conflict of interest.

\section*{Data Availability Statement}
The original contributions presented in the study are included in the article/supplementary material, and further inquiries can be directed to the corresponding author.

\appendix

\section{Wind farm multivariate power model}\label{power}

The wind farm parameters include freedom wind speed $u_{\infty}$, wind direction, ambient turbulence intensity $I_0$, wind shear coefficient, wind farm scale and WTs' coordinates. Assume all turbines are on the same horizontal level; the coordinates of all WTs in the field horizontal slice at $z=h_0$ are defined as ${X = \{X_1,...,X_n,...,X_N \},Y = \{Y_1,...,Y_n,...,Y_N \}}$. The WT parameters include rotor plane diameter $D$, rotor plane area $A$ and hub height $h_0$. These parameters are all constant.
 
The WT's yaw angle $\gamma$, tilt angle $\beta$ and axial induction factor $\alpha$ are defined as variable input in the condition.

\begin{figure}[h]
\centering
\includegraphics[width=0.9\textwidth]{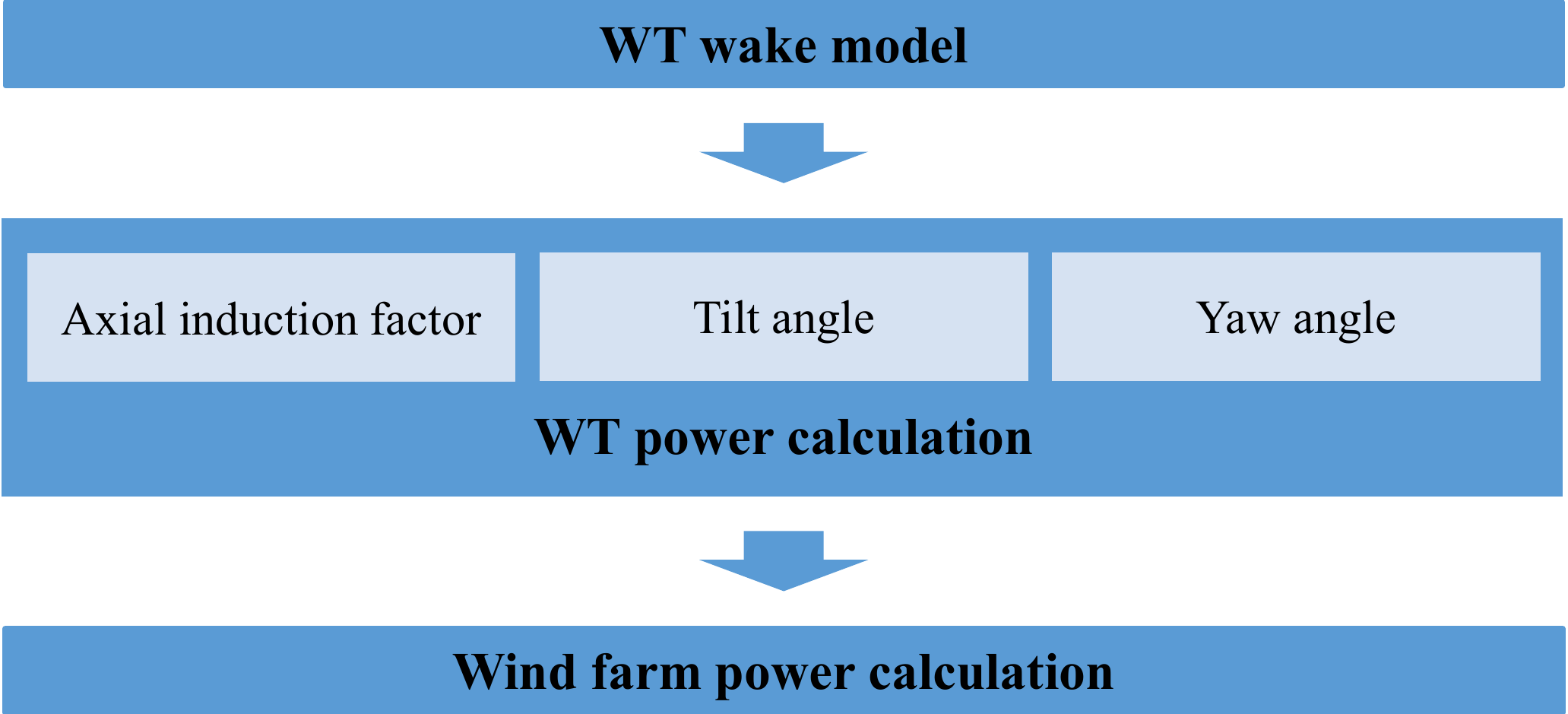}
\caption{Frame of wind farm multivariate power model.}
\label{fig:frame}
\end{figure}

The main structure of the wind farm multivariate power model is shown in Figure~\ref{fig:frame}.

\subsection{WT wake model}

The WT wake model is designed based on the Gauss wake model\citep{bastankhah2016experimental}. Discuss the wake region of the $n$th WT, located at $\{x=X_n,y=Y_n\}$ in the wind farm horizontal slice $z=h_0$, indicating relating parameters by a subscript $n$.

\begin{figure}[htb]
\centering
\includegraphics[width=1.0\textwidth]{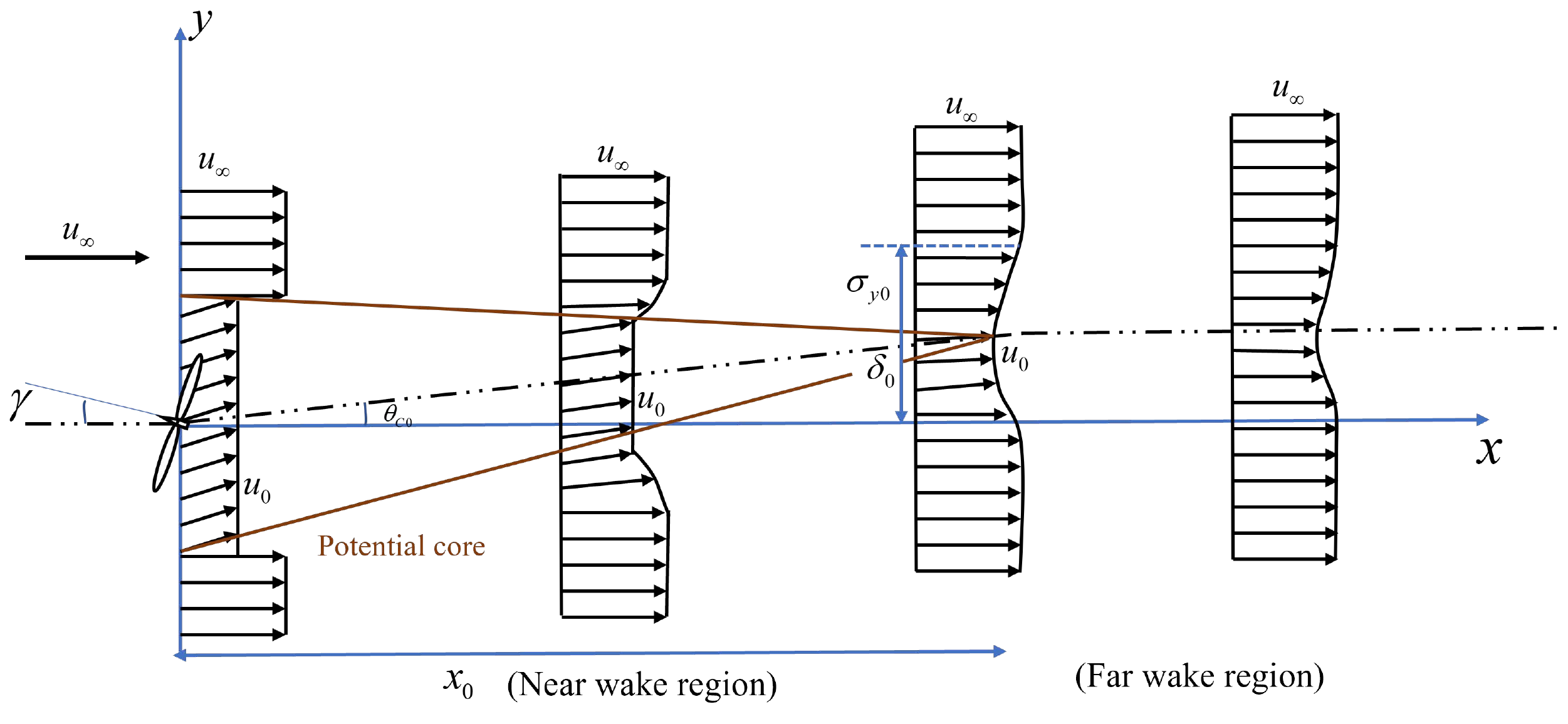}
\caption{Wake region in Gauss wake model.}
\label{fig:wake_region}
\end{figure}

Figure~\ref{fig:wake_region} shows an idealized schematic of the wake of a yawed WT in the Gauss wake model. The potential core, a region with a uniform velocity just behind the WT,  gradually demises as it moves downstream due to its interaction with the surroundings \citep{rajaratnam1976turbulent}. After a certain downwind distance, namely $x_0$, the potential core ultimately disappears, and the velocity profile develops fully into a self-similar Gauss distribution.
\begin{align}
&x_{0,n}=\frac{D_n cos(\gamma_n)(1+\sqrt{1-C_{T,n} cos(\gamma_n)})}{\sqrt{2} (4k_{\alpha} I_n +2k_{\beta} \sqrt{1-C_{T,n}})}\label{Eq:x0},\\
&I_n = \sqrt{I_{0,n}^2+{\Delta I_n}^2} \label{equal:15},\\
&\Delta I_n = 0.9\alpha_n^{0.8}I_{0,n}^{0.1}(\frac{x-X_n}{D_n})^{-0.32}\label{equal:dI}
\end{align}
where ${D_n}$ is the diameter of the $n$th WT rotor plane. ${\gamma_n}$ is the yaw angle of the $n$th WT, $C_{T,n}$ is the thrust coefficient of the $n$th WT. $k_{\alpha}$ and $k_{\beta}$ are parameters used to determine the dependence of potential core length on the $n$th WT's turbulence intensity $I_n$ and axial induction factor $\alpha_n$, defined as 0.58 and 0.077, respectively. Turbulence intensity $I_n$ can be calculated by (\ref{equal:15}). The WT wake flow leads to a substantial increase in turbulence intensity concerning the turbulence level of the incoming atmospheric boundary layer flow, defined as ambient turbulence intensity ${I_{0,n}}$. The added turbulence intensity $\Delta I_n$ is calculated as (\ref{equal:dI}). 

Across the potential core, air velocity remains equal to $u_0$ from $x=0$ to $x_0$ based on the assumption in the Gauss wake model, also called near wake region.
\begin{align}
u_{0,n}=u_{\infty} \sqrt{1-C_{T,n}} \label{Eq:u0},
\end{align}
where, $u_{\infty}$ represents the free-stream velocity velocity shown in Figure~\ref{fig:wake_region}.
 
The wake deflection skew in the near wake region $\theta_{C0,n}$ shown in Figure~\ref{fig:wake_region} is calculated as
\begin{align}
\theta_{C0,n}=\frac{0.3\gamma_n}{cos(\gamma_n)}(1-\sqrt{1-C_{T,n}cos(\gamma_n)})\label{Eq:theta_C0}
\end{align}

The WT far wake expands approximately linearly with $x$ in the streamwise range, therefore, $\sigma_{y,n}$ and $\sigma_{z,n}$ can be estimated by

\begin{align}
&\sigma_{y,n}=k_y{(x-x_{0,n})}+\sigma_{y0,n}, \label{Eq:sigma_y}\\
&\sigma_{z,n}=k_z{(x-x_{0,n})}+\sigma_{z0,n},\label{Eq:sigma_z}\\
&k_y=k_z=0.38I_n+0.004,\label{equal:kyz}
\end{align}

where $\sigma_{y,n}$ and $\sigma_{z,n}$ are different wake widths in the $y$ and $z$ directions in the far wake region for the $n$th WT, as well as $x>x_0$. $k_y$ and $k_z$ are wake growth rates in $y$ and $z$ directions calculated by (\ref{equal:kyz}). $\sigma_{z0,n}$ and $\sigma_{y0,n}$ are initial wake widths in $z$ and $y$ directions, respectively.

\begin{equation}
\begin{aligned}
\sigma_{z0,n}=\frac{1}{2}D\sqrt{\frac{u_{R,n}}{u_{\infty}+u_{0,n}}}, \label{Eq:sigma_yz0}\\
\sigma_{y0,n}=\sigma_{z0,n}cos(\gamma_n)cos(\alpha_n),
\end{aligned}
\end{equation}
where $u_{R,n}$ is the wind velocity at the $n$th WT's rotor. In the Gauss wake model, $u_{R,n}$ is calculated as

\begin{align}
u_{R,n}=u_{\infty}\frac{C_{T,n}cos(\gamma_n)}{2(1-\sqrt{1-C_{T,n}cos(\gamma_n)})}. \label{Eq:uR}
\end{align}

Considering the influence of tilt angle $\beta$, the (\ref{Eq:uR}) in this paper is replaced by

\begin{align}
u_{R,n}=u_{\infty}\frac{C_{T,n}cos(\gamma_n)cos(\beta_n)}{2(1-\sqrt{1-C_{T,n}cos(\gamma_n)cos(\beta_n)})}. \label{Eq:uR_alter}
\end{align}

Moreover, $\delta_n$ is wake deflection at each downwind location of the $n$th WT. The wake deflection of the $n$th WT at the near wake region $\delta_{0,n}=\theta_{C0,n}(x-X_n)$ when $X_n<x<X_n+x_{0,n}$. When $x>X_n+x_{0,n}$, the wake deflection at the far wake region of the $n$th WT $\delta_{n}$ can be determined by

\begin{align}
&\delta_{n} = \frac{\theta_{C0,n}E_{0,n}}{5.2}\sqrt{\frac{\sigma_{z0,n}\sigma_{y0,n}}{k_yk_zM_{0,n}}}ln(\frac{(1.6+\sqrt{M_{0,n}})(1.6\sqrt{1.6\frac{\sigma_{y,1}\sigma_{z,1}}{\sigma_{y0,n}\sigma_{z0,1}}}-\sqrt{M_{0,n}})}{(1.6-\sqrt{M_{0,n}})(1.6\sqrt{\frac{\sigma_{y,n}\sigma_{z,n}}{\sigma_{y0,1}\sigma_{z0,i}}}+\sqrt{M_{0,n}})}) \nonumber\\
&\qquad +\delta_0,\label{Eq:delta}
\end{align}
where $C_{0,n}=1-u_{0,n}/u_{\infty}$ means speed drop ratio, $M_{0,n}=C_{0,n}(2-C_{0,n})$, $E_{0,n}=C_{0,n}^2-3e^{\frac{1}{12}}C_{0,i}+3e^{\frac{1}{3}}$. Although theoretically wake will extend to infinity, for the convenience of calculation, this paper limits the above equation in $x<X_n+15D_n$.

The $n$th WT's wake velocity $u_n(x,y,z)$ and wake deflection skew $\theta_n(x,y,z)$ can be represented by self-similar Gaussian distribution from $x>x_0$, where the downstream WT is always located.

\begin{equation}
\begin{aligned}
\frac{u_n(x,y,z)} {u_{\infty}} = 1- C_n e^{- \frac{y-\delta_n}{2\sigma_y^2}} e^{-\frac{(z-h_0)^2}{2\sigma^2_{z,n}}},\\
\frac{\theta_n(x,y,z)}{\theta_{max}}=e^{-\frac{(y-\delta_n+\sigma_{y,n})^2}{2\sigma_{y,n}^2}} e^{-\frac{(z-h_0)^2}{2\sigma_{z,n}^2}},\label{Eq:u}
\end{aligned}
\end{equation}
where $C_n$ is the velocity deficit at the wake center, defined as $C_n=1-\sqrt{1-\frac{(\sigma_{y0,n}\sigma_{z0,n})M_0}{(\sigma_y\sigma_z)}}$. The maximum wake deflection skew is $\theta_{max}$.

\subsection{WT power calculation}

The (\ref{Eq:u}) calculates velocity in the WT wake region. Define the effective velocity of the $n$th WT as $u_n$, which will be discussed later. The empirical formula to estimate the power associated with the yaw angle $\gamma_n$ is:

\begin{align}
P(\gamma_n) = P_0cos^3(\gamma_n),
\end{align}
where $P_0$ is turbine-rated power without misalignment. However, the empirical parameter $3$ makes substantial errors depending on the incident winds in the atmospheric boundary layer. An alternative method is proposed based on India's multi-month wind farm experiment. Based on static experimental data\citep{howland2022collective}, the power-yaw curve is fitted to predict the influence of yaw angle on power.

Define the yaw coefficient as

\begin{figure}[h]
\centering
\includegraphics[width=0.5\textwidth]{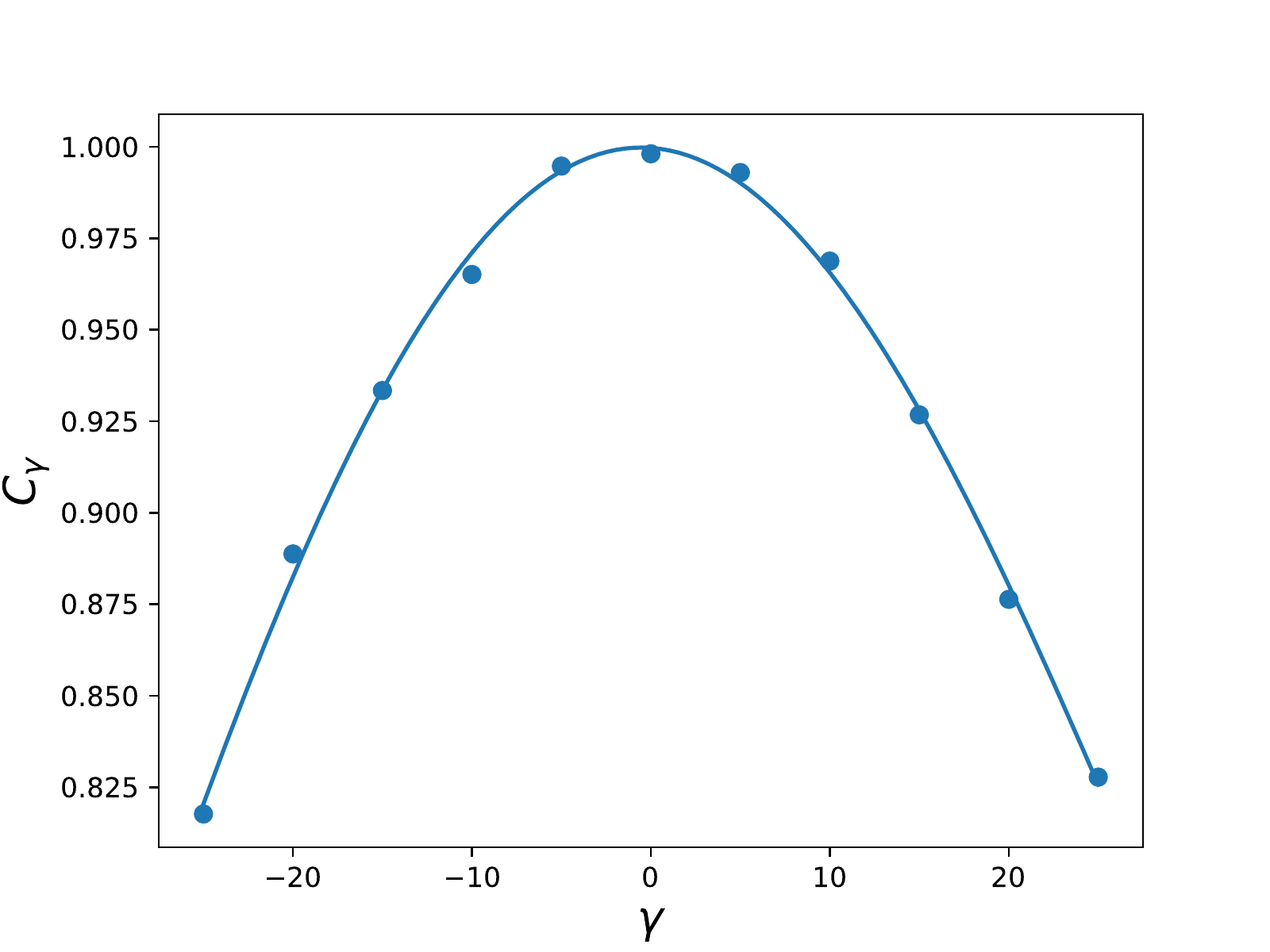}
\caption{Fitted curve for yaw coefficient and yaw.}
\end{figure}

\begin{align}
{C_{\gamma}}(\gamma_n)=(-0.0003{{\gamma_n}^{3}}-0.000025{{\gamma_n}^{2}}\text{+}0.997\gamma_n)\label{Cgamma}.
\end{align}

The influence of tilt angle is also considered in this paper. Like the yaw angle, vertical wake redirection is obtained by changing the tilt angle. The tilt coefficient is defined as

\begin{equation}
\begin{split}
{C_{\beta }(\beta_n)}=&-0.185+0.285\cos (0.105\beta_n)+0.014\sin (0.105\beta_n) \\
&-0.144\cos (0.209\beta_n)-0.017\sin (0.209\beta_n) \\
& +0.06\cos (0.314\beta_n)+0.011\sin (0.314\beta_n) \\
& -0.016\cos (0.419\beta_n)-0.005\sin (0.419\beta_n).
\end{split}\label{Cbeta}
\end{equation}

\begin{figure}[h]
\centering
\includegraphics[width=0.5\textwidth]{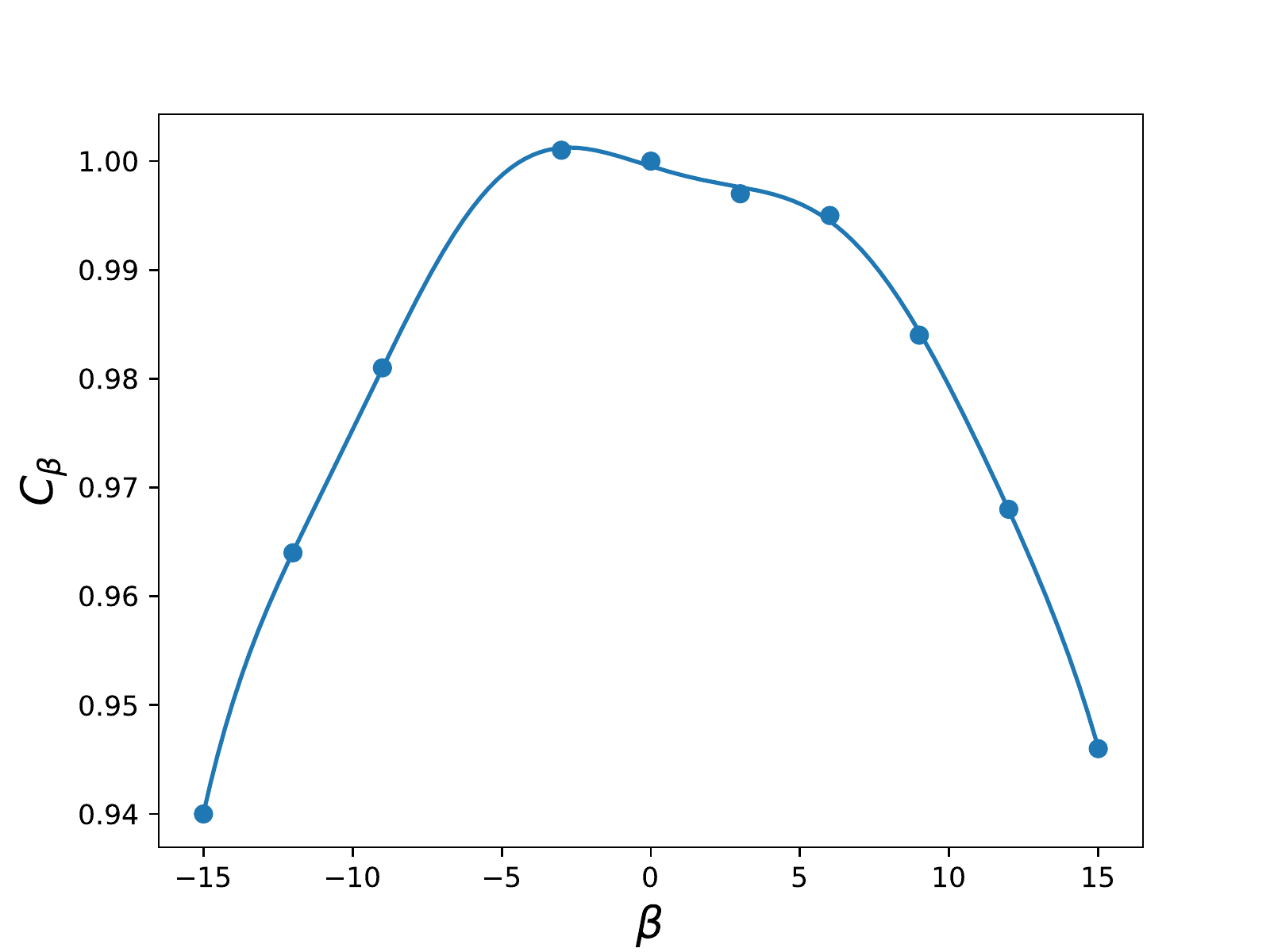}
\caption{Fitted curve for tilt coefficient and tilt angle.}
\label{fig:beta}
\end{figure}

The (\ref{Cbeta}) is obtained by fitting the data proposed by \citet{wang2020influence}, shown in Figure~\ref{fig:beta}.

In conclusion, the $n$th WT power is calculated as

\begin{align}
P_n = \frac{1}{2} C_p(\alpha_n) C_{\gamma}(\gamma_n) C_{\beta}(\beta_n)\rho A_n u_n^3 \label{Eq:P},
\end{align}
where, $\rho$ is the air density, and $A_n$ is the $n$th WT's rotor area, $A_n=\frac{1}{4}\pi {{D_n}^{2}}$. $u_n$ is the wind speed in front of $n$th WT, denoting the effective velocity at the $n$th WT's location, which wolud be discussed after. The WT power coefficient $C_p(\alpha_n)$, is a function of axial induction factor $\alpha_n$, $C_p(\alpha_n)=4{\alpha_n}(1-{\alpha_n})^2$. As shown in (\ref{Eq:P}), yaw angle, tilt angle, and axial induction factor affect the WT power output.

\subsection{Wind farm power calculation}

\begin{figure}[h]
\centering
\includegraphics[width=0.9\textwidth]{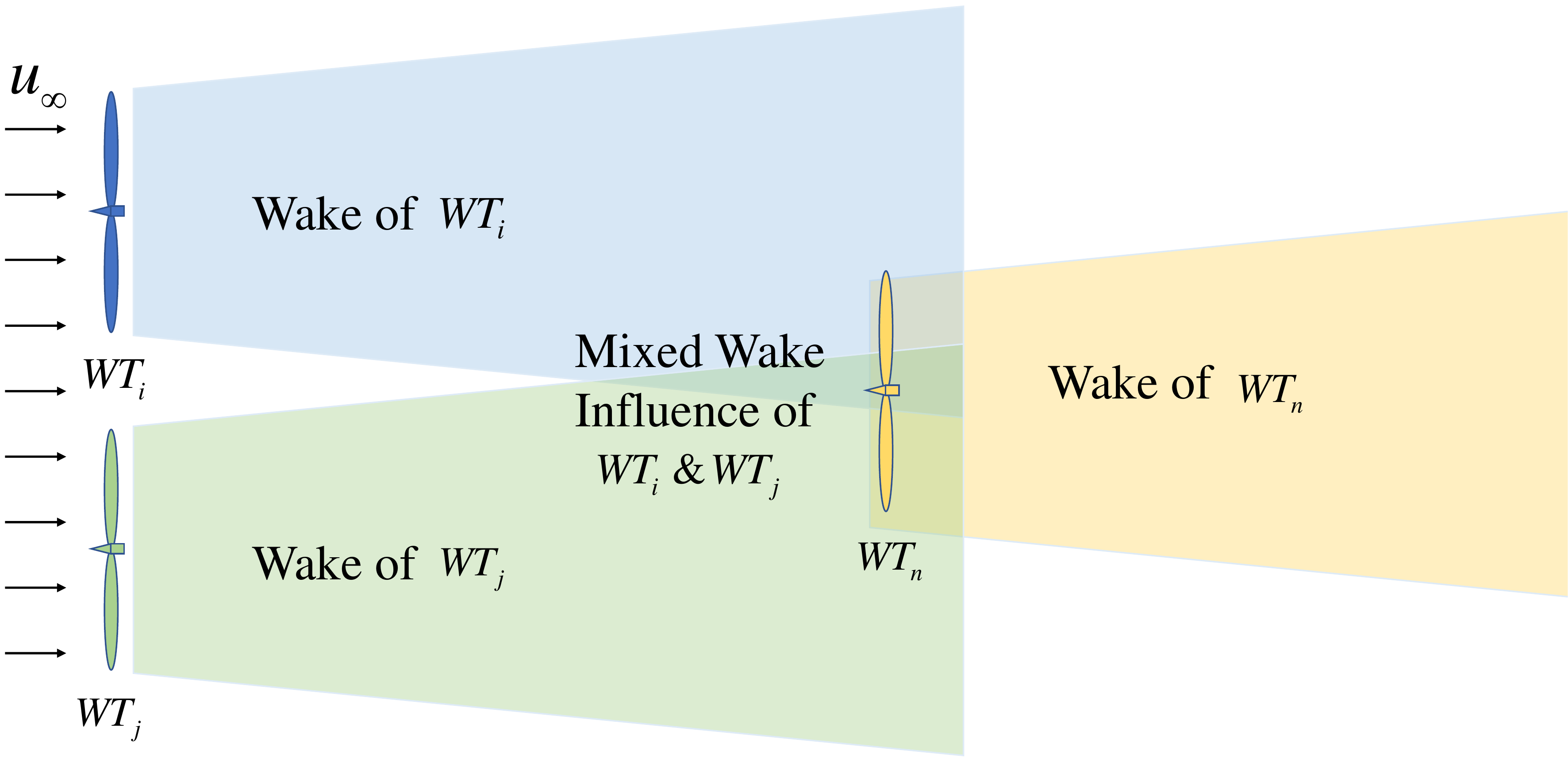}
\caption{Mixed wake influencd.}
\label{fig:mixed_wake}
\end{figure}

As shown in Figure~\ref{fig:mixed_wake}, the the $n$th WT in the overlapping area of wake region of WT $i$ and WT $j$, the effective velocity of the $n$th WT $u_n$ is calculated as
\begin{align}
u_n=u_{\infty}-(u_{\infty}-u_i(X_n,Y_n,h_0))-(u_{\infty}-u_j(X_n,Y_n,h_0)), \label{Eq:effc_v_k}
\end{align}
where $u_{i/j}(X_n,Y_n,h_0)$ is calculated by (\ref{Eq:u}), means the wake velocity of the $i/j$th WT at the $n$th WT's location.

\begin{figure}[h]
\centering
\includegraphics[width=0.9\textwidth]{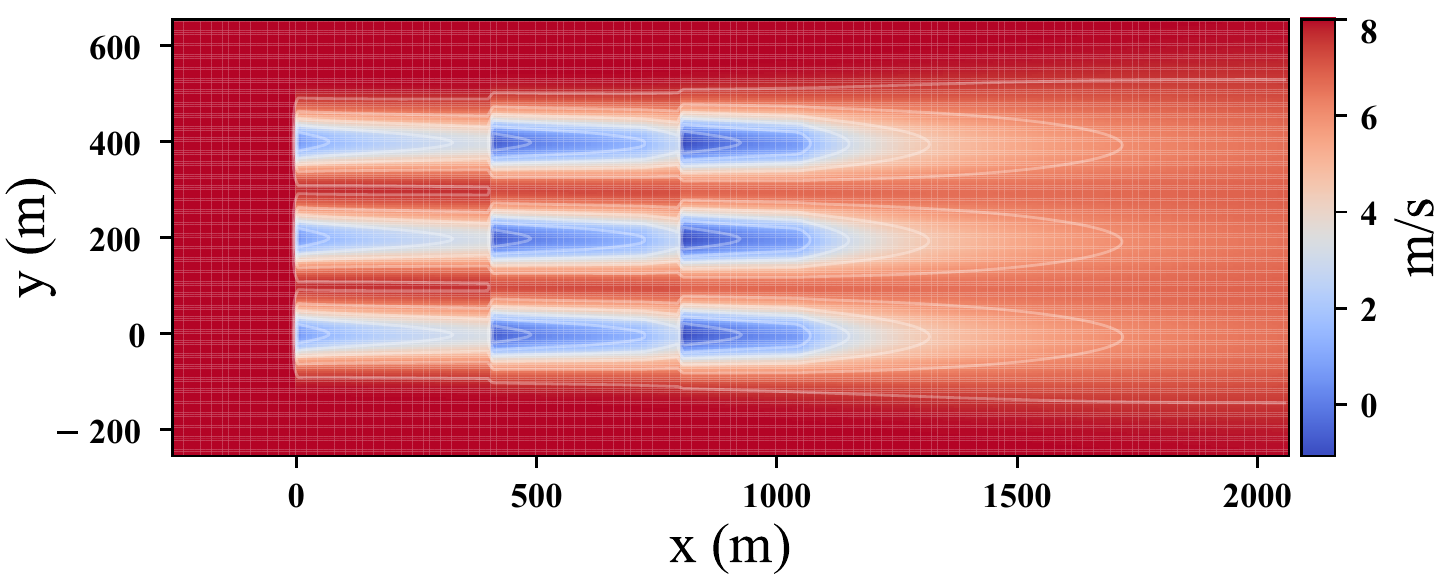}
\caption{The flow field in the nine-turbine wind farm.}
\label{fig:turbine_align}
\end{figure}

As shown in Figure~\ref{fig:turbine_align},  a nine-turbine wind farm with is aligned as wind direction. The downstream WT gets a lower wake velocity because of the wake influence from the upstream WTs. Therefore, the effective velocity at the $i$th WT's position in a wind farm can be calculated as

\begin{equation}
u_n = u_{\infty} - \sum_{i=1,i\neq n}^N (u_{\infty}-u_i(X_n,Y_n,h_0)).\label{Eq:effc_v_k_wf}
\end{equation}

\begin{figure}[h]
\centering
\includegraphics[width=0.9\textwidth]{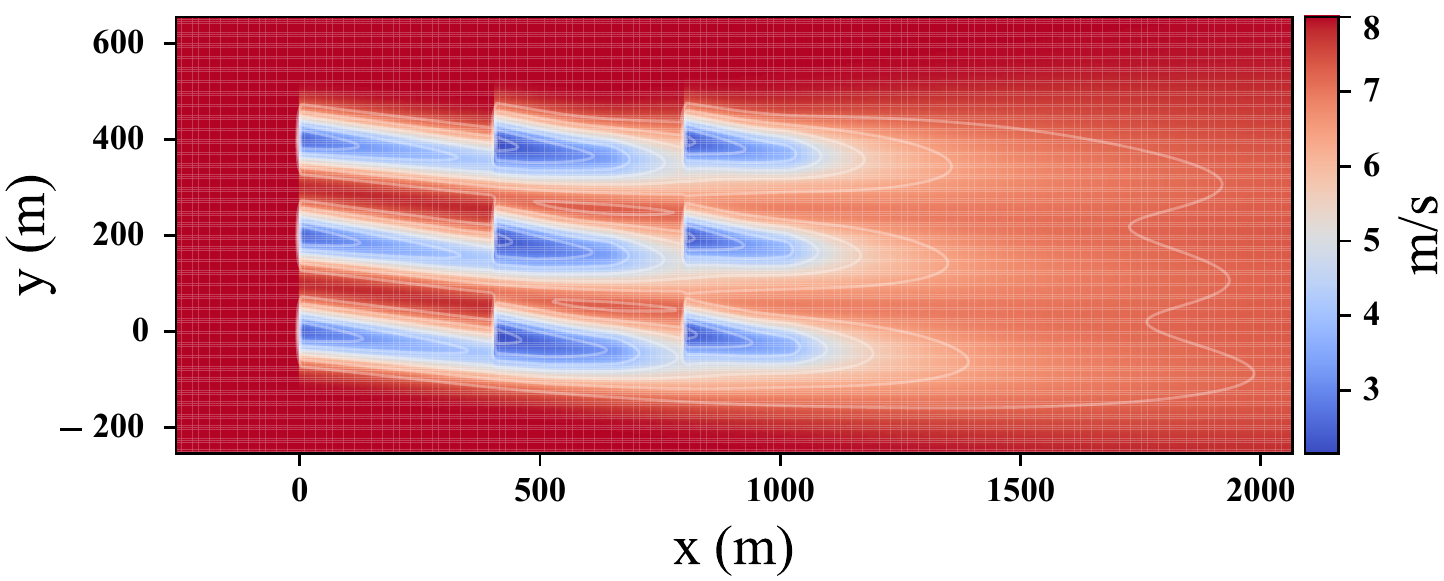}
\caption{The flow field in the nine-turbine wind farm when WTs are misaligned.}
\label{fig:turbine_misalign}
\end{figure}

Moreover, Figure~\ref{fig:turbine_misalign} shows a simulation picture when WTs in Figure~\ref{fig:turbine_align} are misaligned by turning the yaw angle. As shown in Figure~\ref{fig:turbine_misalign}, turning the yaw angle could effectively mitigate upstream WTs' wake influence. Effective velocity at $n$th WT in the wind farm is gotten by (\ref{Eq:effc_v_k_wf}).

In conclusion, effective velocity at each WT, $\{u_1,...,u_n,..,u_N\}$, can be calculated as (\ref{Eq:effc_v_k_wf}). Introduce effective velocity $\{u_1,...,u_n,...,u_N\}$ into (\ref{Eq:P}), and then the power output of each WT, $\{P_1,...,P_n,...P_N\}$, can be gotten. Therefore, the wind farm power output $P_{WF}$ is 

\begin{equation}
P_{WF} =  \sum_{n=1}^N{P_n} .\label{Eq:pwf}
\end{equation}

\section{Layout and division of wind farms}\label{layout}
Layouts of different wind farms are presented in Table~\ref{tab:dlwf}. The WTs in each wind farm are divided into several WTAs, whose numbers and corresponding WT IDs are displayed in the final column of Table~\ref{tab:dlwf}.

\begin{table*}[htbp]
  \centering
  \caption{Layout and division of wind farms}
 \label{tab:dlwf}
  \begin{tabular}{ | c | c| c | c |}
    \hline
    \makecell{Number\\ of WTs} & \makecell{Number\\ of WTAs} & Layout  & WTs and WTAs\\ \hline
     13
& 4
&\begin{minipage}[b]{0.5\columnwidth}
		\centering
		\raisebox{-.5\height}{\includegraphics[width=0.82\textwidth]{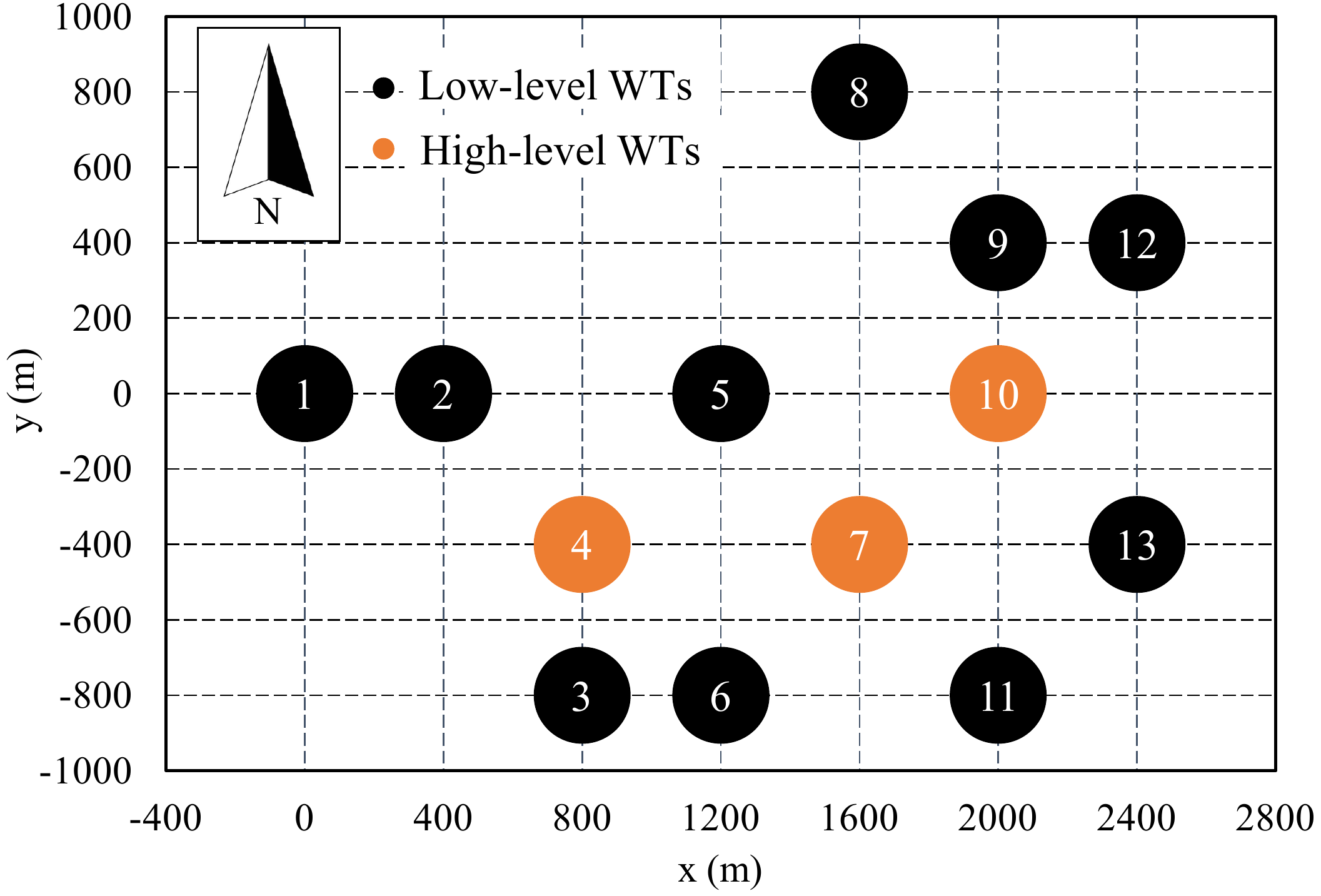}}
	\end{minipage}
    & \makecell{$WTA1=\left\{1,2,3,4\right\}$, \\ $WTA2=\left\{4,5,6,7\right\}$,\\ $WTA3=\left\{7,8,9,10\right\}$,\\ $WTA4=\left\{10,11,12,13\right\}$}
    \\ \hline
     16
& 5
&\begin{minipage}[b]{0.5\columnwidth}
		\centering
		\raisebox{-.5\height}{\includegraphics[width=0.78\textwidth]{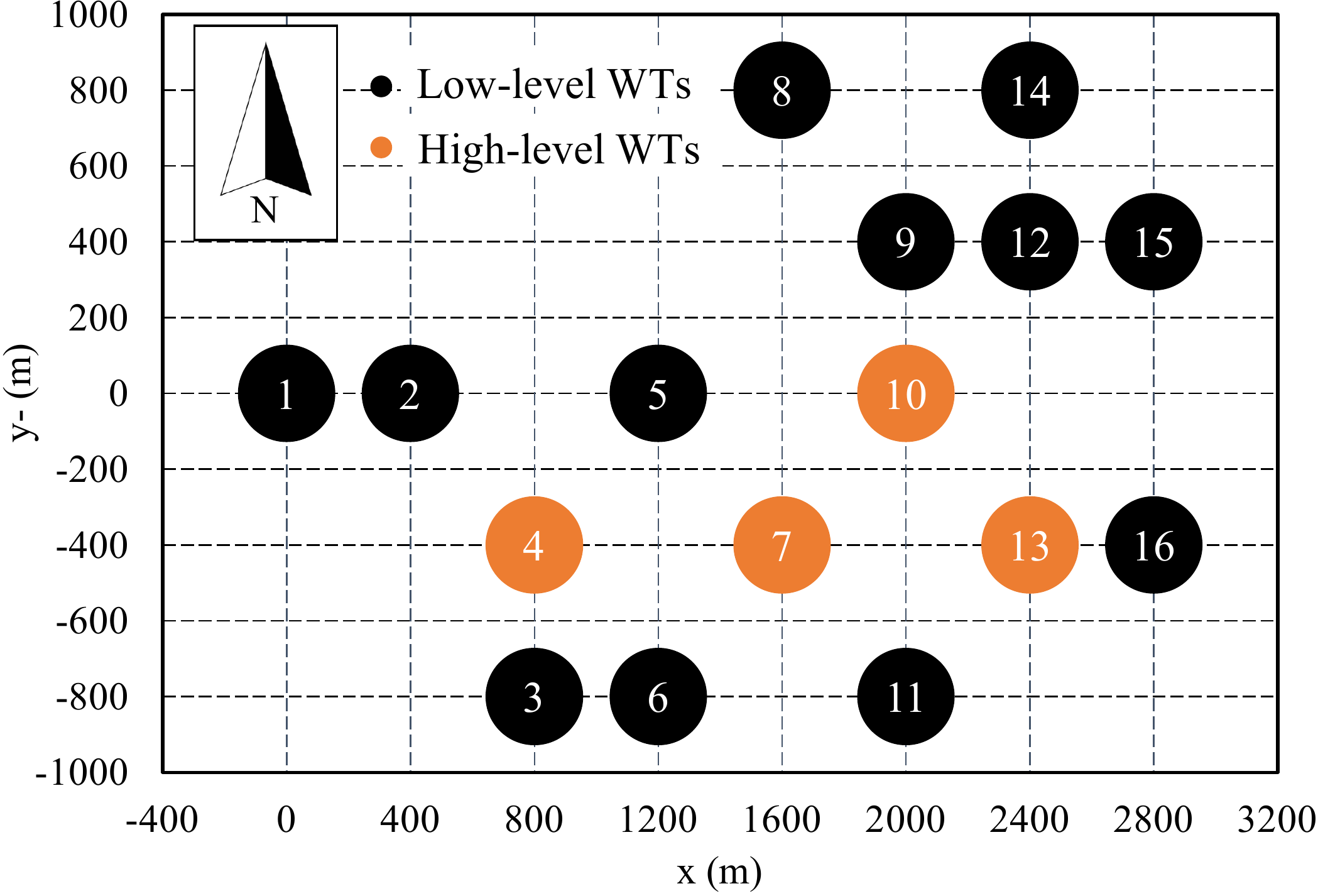}}
	\end{minipage}
    & \makecell{$WTA1=\left\{1,2,3,4\right\}$, \\ $WTA2=\left\{4,5,6,7\right\}$,\\ $WTA3=\left\{7,8,9,10\right\}$,\\ $WTA4=\left\{10,11,12,13\right\}$,\\ $WTA5=\left\{13,14,15,16\right\}$}
    \\ \hline
     19
& 6
&\begin{minipage}[b]{0.5\columnwidth}
		\centering
		\raisebox{-.5\height}{\includegraphics[width=0.9\textwidth]{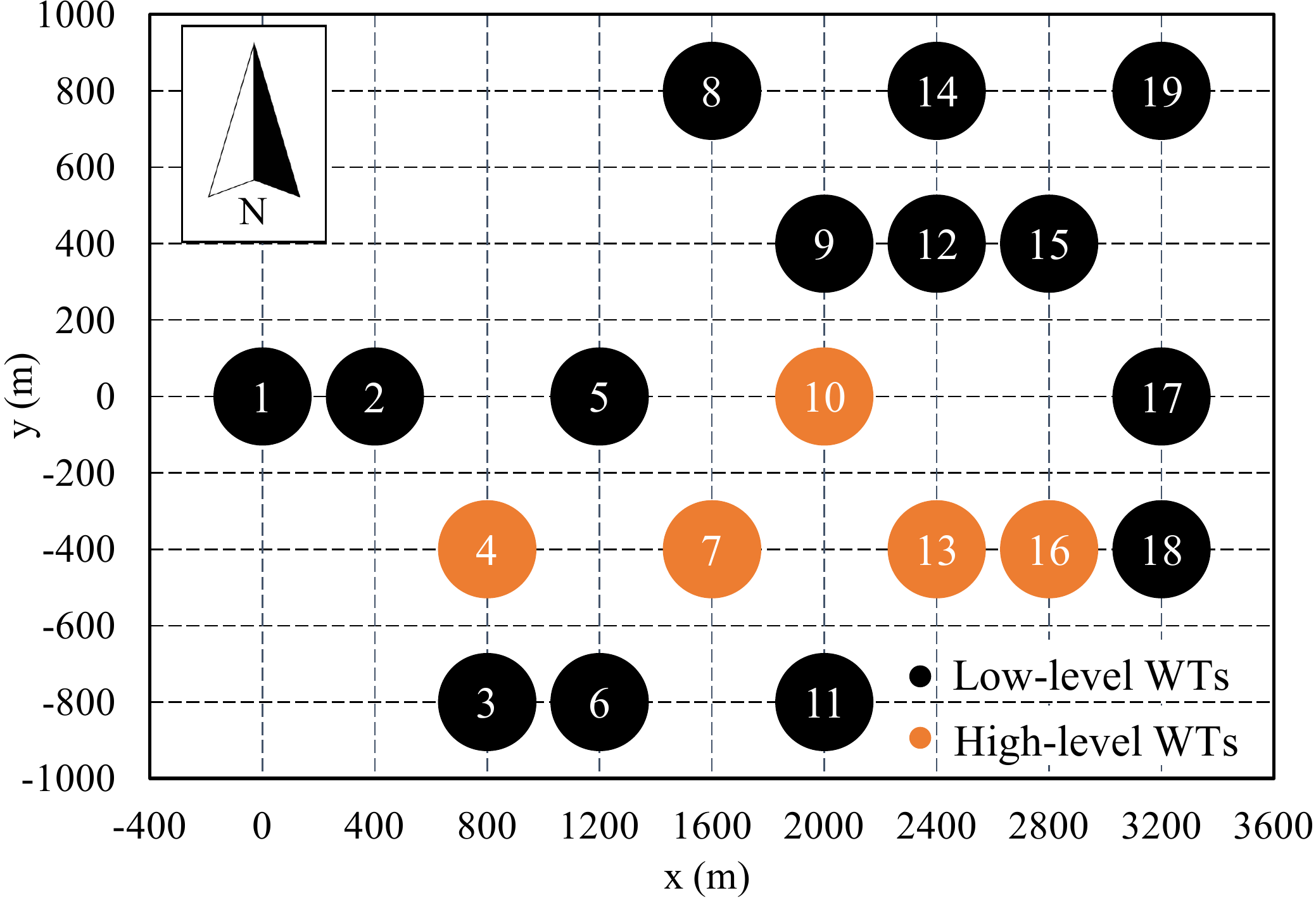}}
	\end{minipage}
    & \makecell{$WTA1=\left\{1,2,3,4\right\}$, \\ $WTA2=\left\{4,5,6,7\right\}$,\\ $WTA3=\left\{7,8,9,10\right\}$,\\ $WTA4=\left\{10,11,12,13\right\}$,\\ $WTA5=\left\{13,14,15,16\right\}$,\\ $WTA6=\left\{16,17,18,19\right\}$}
    \\ \hline
22
& 7
&\begin{minipage}[b]{0.5\columnwidth}
		\centering
		\raisebox{-.5\height}{\includegraphics[width=1.0\textwidth]{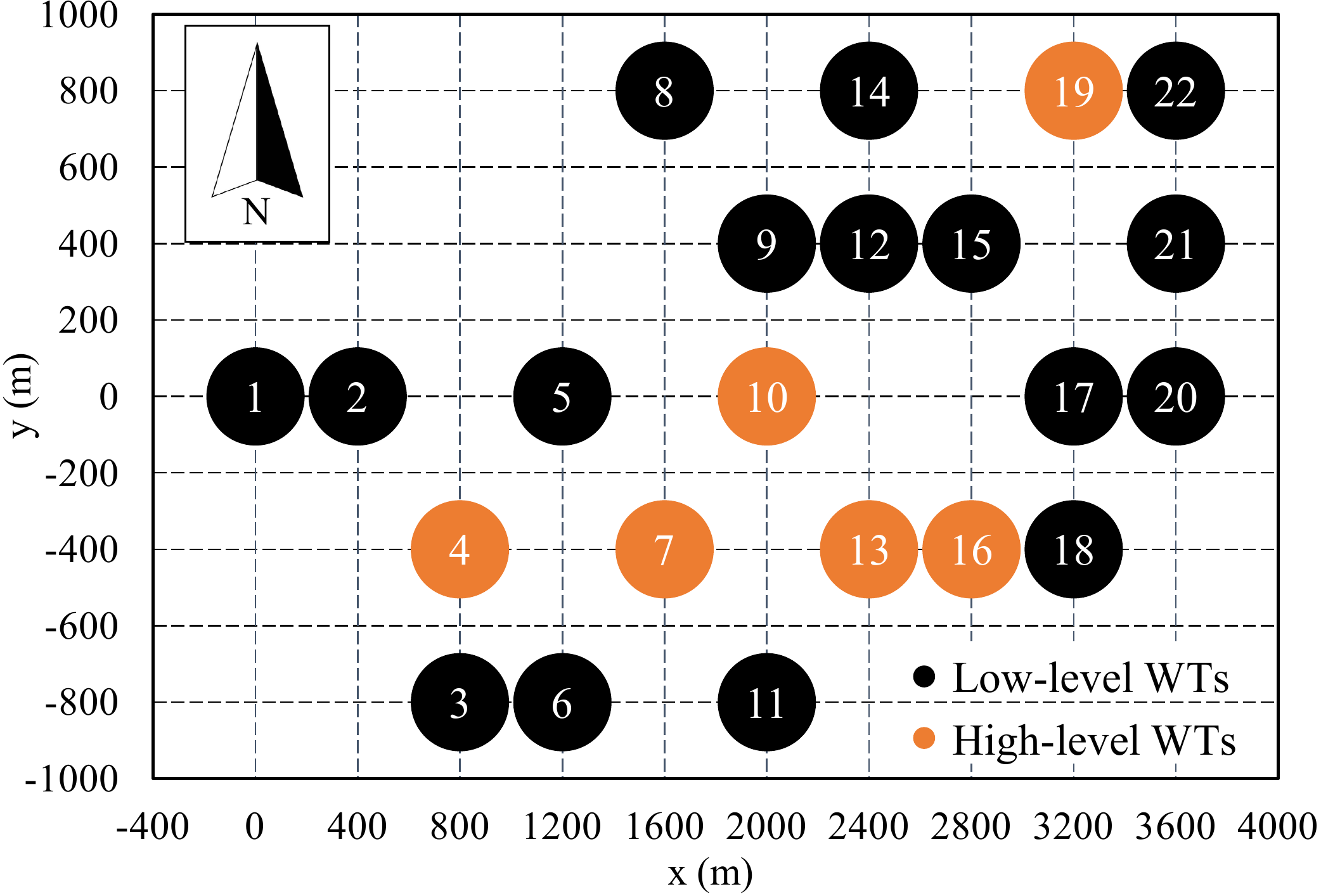}}
	\end{minipage}
    & \makecell{$WTA1=\left\{1,2,3,4\right\}$, \\ $WTA2=\left\{4,5,6,7\right\}$,\\ $WTA3=\left\{7,8,9,10\right\}$,\\ $WTA4=\left\{10,11,12,13\right\}$,\\ $WTA5=\left\{13,14,15,16\right\}$,\\ $WTA6=\left\{16,17,18,19\right\}$,\\ $WTA7=\left\{19,20,21,22\right\}$}
    \\ \hline
  \end{tabular}
\end{table*}

\bibliographystyle{elsarticle-num-names} 
\bibliography{ref}

\end{document}